\documentclass[pre,aps,twocolumn,superscriptaddress]{revtex4}
\pdfoutput=1
\usepackage{amssymb}
\usepackage{epsfig}
\usepackage{amsmath}
\usepackage{subfigure}
\usepackage{graphicx}
\usepackage{textcomp}
\usepackage{url}
\usepackage{float}
\usepackage{color}
\usepackage{cases}
\usepackage[normalem]{ulem}

\usepackage[titletoc,title]{appendix}

\usepackage[colorlinks=true, urlcolor=blue, anchorcolor=blue, citecolor=blue,filecolor=blue,linkcolor=blue,menucolor=blue
]{hyperref}

\begin{document}
\title{Geometrical optics of large deviations of fractional Brownian motion}

\author{Baruch Meerson}
\affiliation{Racah Institute of Physics, Hebrew University of
Jerusalem, Jerusalem 91904, Israel}
\author{Gleb Oshanin}
\affiliation{Sorbonne Universit\'{e}, CNRS, Laboratoire de Physique Th\'{e}orique de la Mati\`{e}re Condens\'{e}e (UMR CNRS 7600), 4 Place Jussieu, 75252
Paris Cedex 05, France}

\begin{abstract}
It has been shown recently that the optimal fluctuation method -- essentially geometrical optics -- provides a
valuable insight into large deviations of Brownian motion. Here we extend the geometrical optics
formalism to two-sided, $-\infty<t<\infty$, fractional Brownian motion (fBM) on the line, which is ``pushed" to a large deviation regime by imposed constraints. We test the formalism on three examples where exact solutions are available: the two- and three-point probability distributions of the fBm and the distribution of the area under the fBm on a specified time interval.
Then we apply the formalism to several previously unsolved problems by evaluating large-deviation tails of the following distributions: (i) of the first-passage time, (ii) of the maximum of, and (iii) of the area under, fractional Brownian bridge and fractional Brownian excursion, and (iv) of the first-passage area distribution of the fBm. An intrinsic part of a geometrical optics calculation is determination of the optimal path -- the most likely
realization of the process which dominates the probability distribution of the conditioned process.  Due to the non-Markovian nature of the fBm, the optimal paths of a fBm, subject to constraints on a finite interval $0<t\leq T$, involve both the past $-\infty<t<0$ and 
the future $T<t<\infty$.

\end{abstract}

\maketitle
\nopagebreak

\section{Introduction}
\label{intro}

Trajectories of a stochastic process, ``pushed" into a large-deviation regime by imposed constraints, look quite differently from the typical trajectories of the  unconstrained process. Apart from a valuable intuitive insight into the physics of constrained stochastic processes, the focus on large-deviation trajectories can be beneficial in terms of quantitative results. Indeed,  the knowledge of the most   probable trajectory of such a constrained process often suffices for the evaluation of the probability distributions of the constrained process. This is the essence of the optimal fluctuation method, which goes back to Refs. \cite{Halperin,Langer,Lifshitz}, and which has found numerous application in different areas of physics. Recently, the optimal fluctuation method has been developed and applied in a number of studies  of large deviations of Brownian motion with various constraints \cite{M2019a,SM2019,M2019b,MS2019,MM2020a,Agranovetal,M2020}.
In the context of Brownian motion the OFM essentially becomes geometrical optics, where
the optimal path is a geodesic subject to imposed constraints.

In this work we extend the geometrical optics approach to large deviations of  \emph{fractional} Brownian motion (fBm): a generalization of the standard Brownian motion which is non-Markovian,  but still  keeps the important properties of dynamical scale invariance, stationarity of the increment and Gaussianity \cite{Kolmogorov,Mandelbrot}. There is a variety of physical processes that have been successfully modeled as fBm:
anomalous diffusion in different media \cite{Bouchaud,weiss,weiss2,weber,ralf}, diffusion of a tagged monomer inside a polymer \cite{Walter,Amitai}, translocation of a polymer through a pore \cite{Amitai,Zoia,Dubbeldam,Palyulin}, single-file
diffusion in ion channels \cite{Kukla,Wei}, \textit{etc.}

Large-deviation statistics of fBm were previously studied in the context of single-file diffusion \cite{Sadhu1,Sadhu2,Sadhu3}. The following circumstance provide an ample motivation for continuing studying large deviations of fBm.  The effective particle motion in a multitude of  living and non-living systems can be described by anomalous diffusion consistent with the fBm model  \cite{weiss,weiss2,weber,ralf}. In living systems, large deviations of particle transport appear naturally when multiple agents (for example, molecules or sperm cells) randomly search for an immobile target site (a receptor or the oocyte), and the reaction occurs upon the first arrival of the first among the agents, see Ref. \cite{Schussreview} and references therein. The arrival of the first particle out of many is necessarily unusually fast. The particle trajectory in this case is very different from a typical one and may be amenable to a geometrical optics approach.

This fact has been recognized for the normal diffusion \cite{Schussreview},  and it gave rise to the development of geometrical optics of Brownian motion. However, its practical recognition in the context of anomalous diffusion, modeled by the fBm, poses significant challenges -- both conceptual and technical -- because of the intrinsic non-Markovian nature of the fBm. In this work we meet these challenges by fully exploiting the Gaussianity of the fBm. We develop a (nonlocal generalization of) the geometrical-optics
formalism for the fBm, which provides a remarkable insight into large deviations of the fBm and gives an accurate evaluation of large-deviation tails of several previously unknown probability distributions.

We introduce the geometrical-optics formalism for fBm in Sec. \ref{fBmoptics}. As the first two tests of the geometrical optics we evaluate, in Sec. \ref{propagator},   the two- and three-point probability distributions of the fBm, and use these results for calculating the tails of two additional related
distributions: of the first passage time to a point, and of the maximum of fractional
Brownian bridge and fractional Brownian excursion  In Sec. \ref{areas} we study distributions of areas under fBm subject to different constraints. First we evaluate, in Sec. \ref{noconstraints}, the area distribution without any constraint. In this simple case the distribution is known exactly, and we use it as an additional  test of the geometrical optics. In Sec. \ref{fBBfBE} we evaluate the large-area tails of the area distributions under a fractional Brownian bridge and fractional Brownian excursion. In
Sec. \ref{fparea} we calculate the distribution of the first-passage area of a fBm. Our main results are summarized and discussed in Sec. \ref{discussion}.

\section{fBm and geometrical optics}
\label{fBmoptics}

Let $x(t)$ denote a realization, such that $x(0)=0$, of a one-dimensional fBm \cite{Kolmogorov,Mandelbrot}:
a Gaussian zero-mean process which generalizes the standard Brownian motion to a family
of ``anomalous diffusion" processes. For the two-sided version of the fBm, the time variable is defined on the entire axis, $-\infty < t < \infty$, and the
covariance of the process is given by the equation
\begin{equation}\label{kappa}
\!\kappa(t,t')\!=\!\langle x(t) x(t')\rangle\!=\!D\left(|t|^{2H}\!+\!|t'|^{2H}\!-\!|t-t'|^{2H}\right),
\end{equation}
where the Hurst exponent $H \in (0,1)$  quantifies the dynamical scale-invariance of the process, see \textit{e.g.} Ref. \cite{Stanley}, and $D=\text{const}>0$ is the coefficient of fractional diffusion.   For the one-sided fBM, $0 \leq t < \infty$, one has
\begin{equation}\label{kappa1}
\kappa_1(t,t')\!=\!\langle x(t) x(t')\rangle\!=\!D\left(t^{2H}+t'^{2H}-|t-t'|^{2H}\right)\,,
\end{equation}
where both $t$ and $t'$ are non-negative. To be specific, we focus here on the two-sided fBm. Extension of our analysis to the one-sided case,  $0 \leq t < \infty$, is straightforward, as we explain in Sec. \ref{discussion}.

\begin{figure}[ht]
\includegraphics[width=0.3\textwidth,clip=]{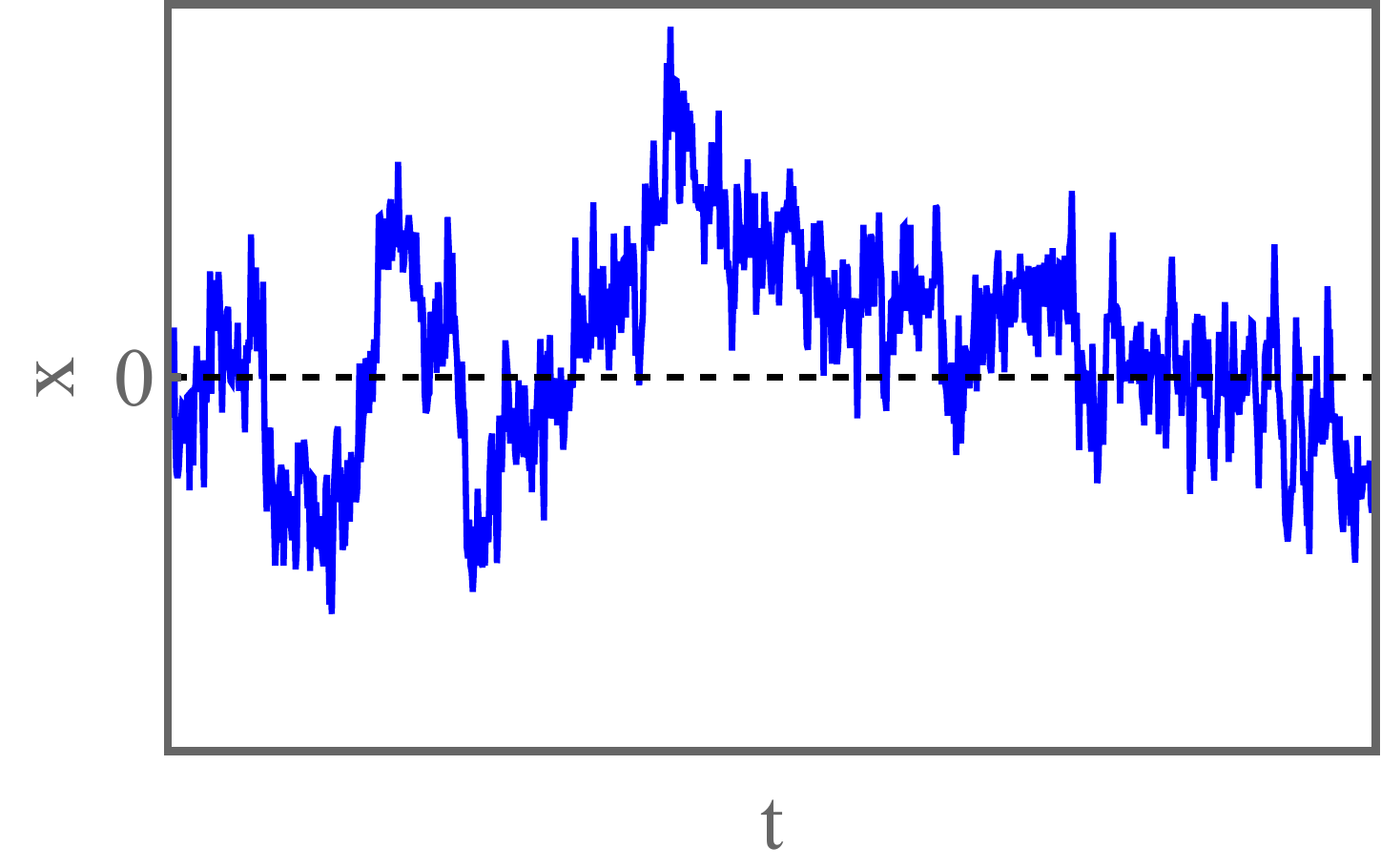}
\includegraphics[width=0.3\textwidth,clip=]{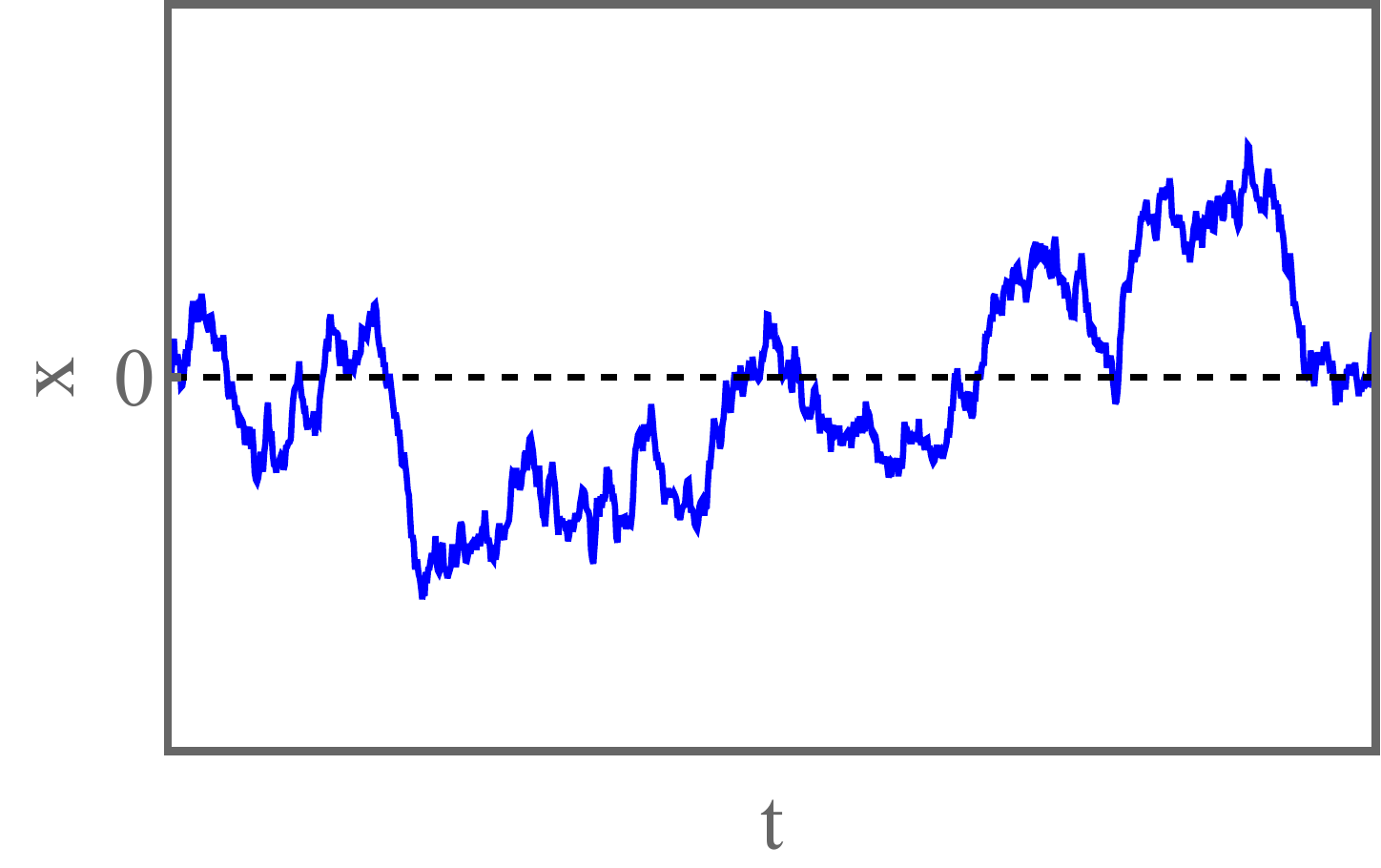}
\includegraphics[width=0.3\textwidth,clip=]{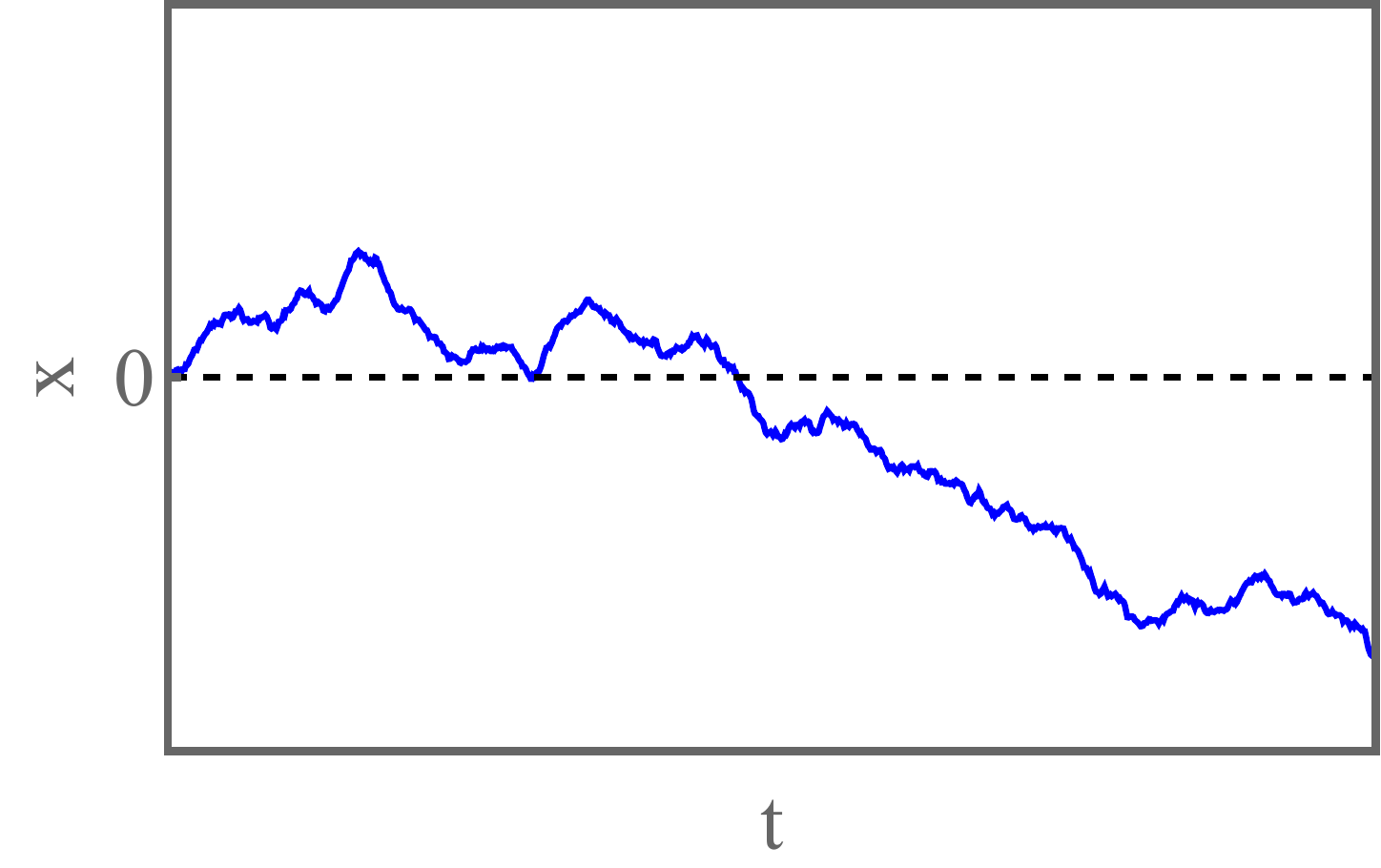}
\caption{Realizations of one-sided fBm for $H=1/4$ (top) and $H=3/4$ (bottom). A realization for $H=1/2$ (middle) corresponds to the standard Brownian motion, $H=1/2$.}
\label{realizations}
\end{figure}

The fBm  describes a family of anomalous diffusions, where
the regimes of $0<H<1/2$ and $1/2<H<1$ correspond to sub-diffusion and super-diffusion, respectively, while the standard diffusion (with $t$ defined on the entire axis) is recovered for $H=1/2$. Figure \ref{realizations} shows some stochastic realizations of fBm for $H=1/4$, $1/2$ (standard Bm) and $3/4$, obtained numerically.

The starting point of the optimal fluctuation method, such as geometrical optics of the fBm, is the probability cost of a given realization $x(t)$. Since the fBm is a Gaussian process, this  probability cost can be written as $\sim \exp\left(-S[x(t)]\right)$ with the action \cite{Zinn}
\begin{equation}\label{actiongeneral}
S[x(t)] = \frac{1}{2} \int_{-\infty}^{\infty} dt \int_{-\infty}^{\infty} dt' K(t,t') x(t) x(t')\,.
\end{equation}
Here $K(t,t')\equiv K(t',t)$ is the inverse kernel, defined by the relation
\begin{equation}\label{inverse}
\int_{-\infty}^{\infty} d \tau \,K(t, \tau) \,\kappa(\tau,t') =\delta(t-t').
\end{equation}
A crucial observation here is that the fBm is a non-Markovian process. Therefore, even when one considers it on a finite interval $0<t< T$, the path integral is also affected by the past, $-\infty<t<0$, and the future, $T<t<\infty$. The latter fact, however, does not violate causality.

In the presence of constraints which push the process into a large-deviation regime, the action
(\ref{actiongeneral}) becomes very large. As a result, the dominating contribution to the probability distribution comes from the \emph{optimal path}: a single deterministic trajectory $x_*(t)$ which minimizes the action functional~(\ref{actiongeneral})  subject to the specified additional constraints. The minimization procedure
leads to a non-local extension of the Euler-Lagrange equation for the optimal path \cite{M2019c,M2022}.
Once the optimal path $x_*(t)$ is determined, one can evaluate the probability distribution of the specific large deviation up to a pre-exponential factor,
\begin{equation}\label{Pgeneral}
- \ln \mathcal{P} \simeq S[x_*(t)]\,,
\end{equation}
by plugging the optimal path into the action functional $S[x(t)]$ given by Eq.~(\ref{actiongeneral}).
We will present an explicit form for the inverse kernel $K(t,t')$ for the two-sided and one-sided fBm elsewhere \cite{a}. In this work we exploit the remarkable fact that, in many geometrical-optics calculations, one does not actually need to know it.

\section{Two- and three-point distributions of the fBm}
\label{propagator}

Two- and three-point distributions belong to fundamental statistics of any stochastic process, so it is natural to start from them.

\subsection{Two-point distribution}

Since $x(t)$ is normally distributed, with the variance $\kappa(t,t) = 2D|t|^{2H}$, the exact
two-point distribition -- the propagator of the fBm -- immediately follows:
\begin{equation}
\label{GreenfBM}
  \mathcal{P}(x=X,t=T)=
  \frac{1}{\sqrt{4\pi D T^{2H}}}\,\,  e^{-\frac{X^2}{4 D T^{2H}}}\,.
\end{equation}
Now we reproduce the propagator (\ref{GreenfBM}), up to the normalization factor, by using geometrical optics of fBm. Assuming a very small $T$ or a very large $X$, we can minimize  the action functional~(\ref{actiongeneral})  over all possible paths $x(t)$ obeying [in addition to the condition $x(0)=0$ which holds automatically] the condition $x(T)=X$.  It is convenient to impose the latter condition through an integral constraint,
\begin{equation}\label{HHH}
\int_{-\infty}^{\infty} x(t) \delta(t-T) dt =X\,.
\end{equation}
Now we can minimize the constrained action
\begin{eqnarray}
 S_{\lambda}[x(t)]&=& \frac{1}{2}\int_{-\infty}^{\infty} dt \left[\int_{-\infty}^{\infty} dt^\prime  K(t,t^\prime) x(t) x(t^\prime) \right. \nonumber\\
  &-& \left. \lambda x(t) \delta(t-T)\right]\,,
\label{functional}
\end{eqnarray}
where $\lambda$ is a Lagrange multiplier to be ultimately expressed through $X$ and $T$. The linear variation
\begin{equation}\label{B90}
 \!\delta s_{\lambda} \!=\! \int_{-\infty}^{\infty} dt \,\delta x(t) \left[\int_{-\infty}^{\infty} dt^{\prime} K(t,t^\prime) x(t^\prime)\!-\! \frac{\lambda}{2} \delta(t-T) \right],
\end{equation}
must vanish for arbitrary $\delta x(t)$, leading to the linear equation
\begin{equation}\label{B100}
\!\int_{-\infty}^{\infty} dt^\prime\,K(t,t^\prime)x(t^\prime) \!= \!\frac{\lambda}{2} \delta(t-T),
\end{equation}
Comparing Eqs.~(\ref{inverse}) and~(\ref{B100}), we immediately obtain the solution up to the still unknown factor $\lambda$:
\begin{equation}\label{xlambda}
x_*(t) =\frac{\lambda}{2} \kappa(t,T)\,.
\end{equation}
Using the condition $x_*(T)=X$, we express $\lambda$ through $X$ and $T$:
\begin{equation}\label{lambda}
\lambda=\frac{2X}{\text{Var}(T)} = \frac{X}{D T^{2H}}\,,
\end{equation}
where $\text{Var}(T) = 2 D T^{2H}$ is the variance of the process. As a result, the optimal path takes the form
\begin{equation}\label{xXT}
x_*(t) = \frac{X}{2D T^{2H}}\kappa(t,T)\,,\quad |t|<\infty\,.
\end{equation}
As one can see, the optimal path of the two-point distribution is proportional to the covariance of the process \cite{M2022}. This simple but important result is quite general and not limited to the fBm: it applies to large deviations of two-point statistics of \emph{any} Gaussian process.

\begin{figure}[ht]
\includegraphics[width=0.3\textwidth,clip=]{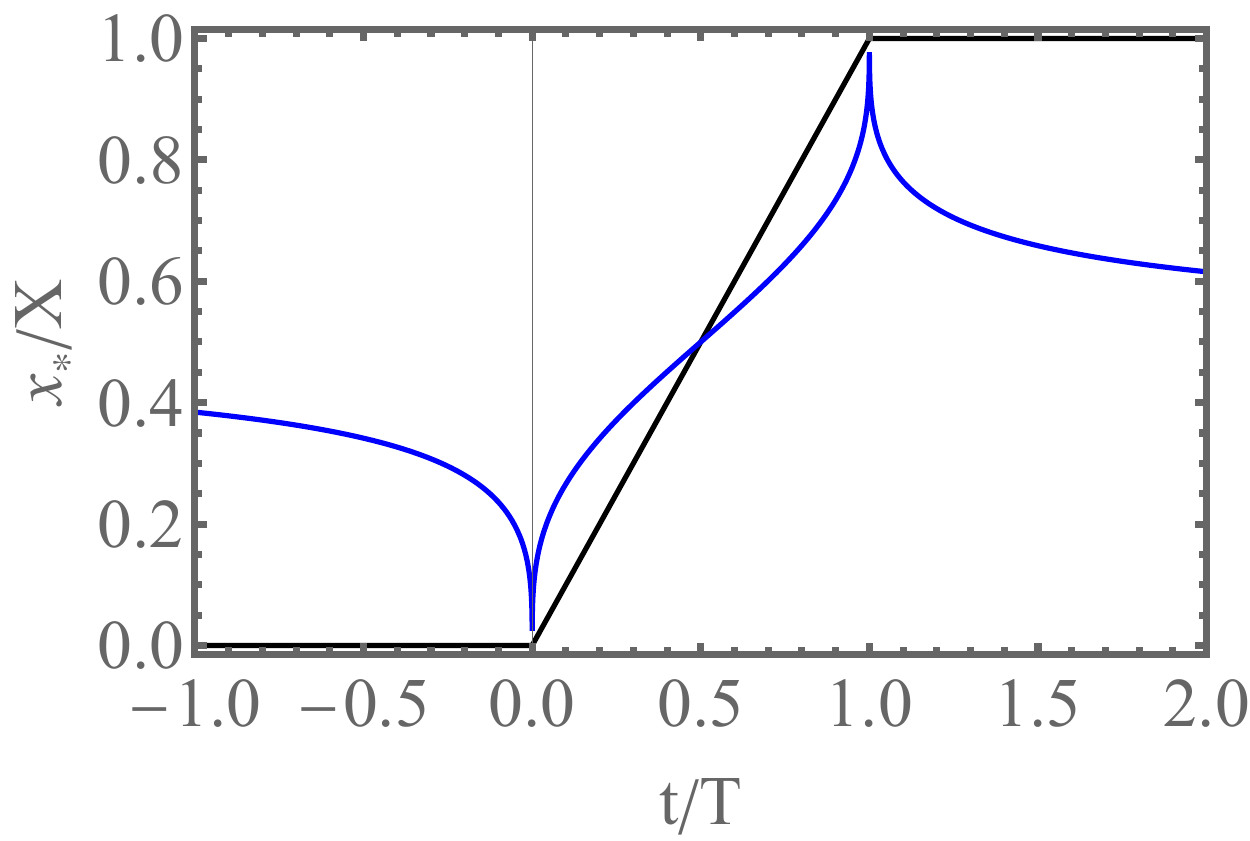}
\includegraphics[width=0.3\textwidth,clip=]{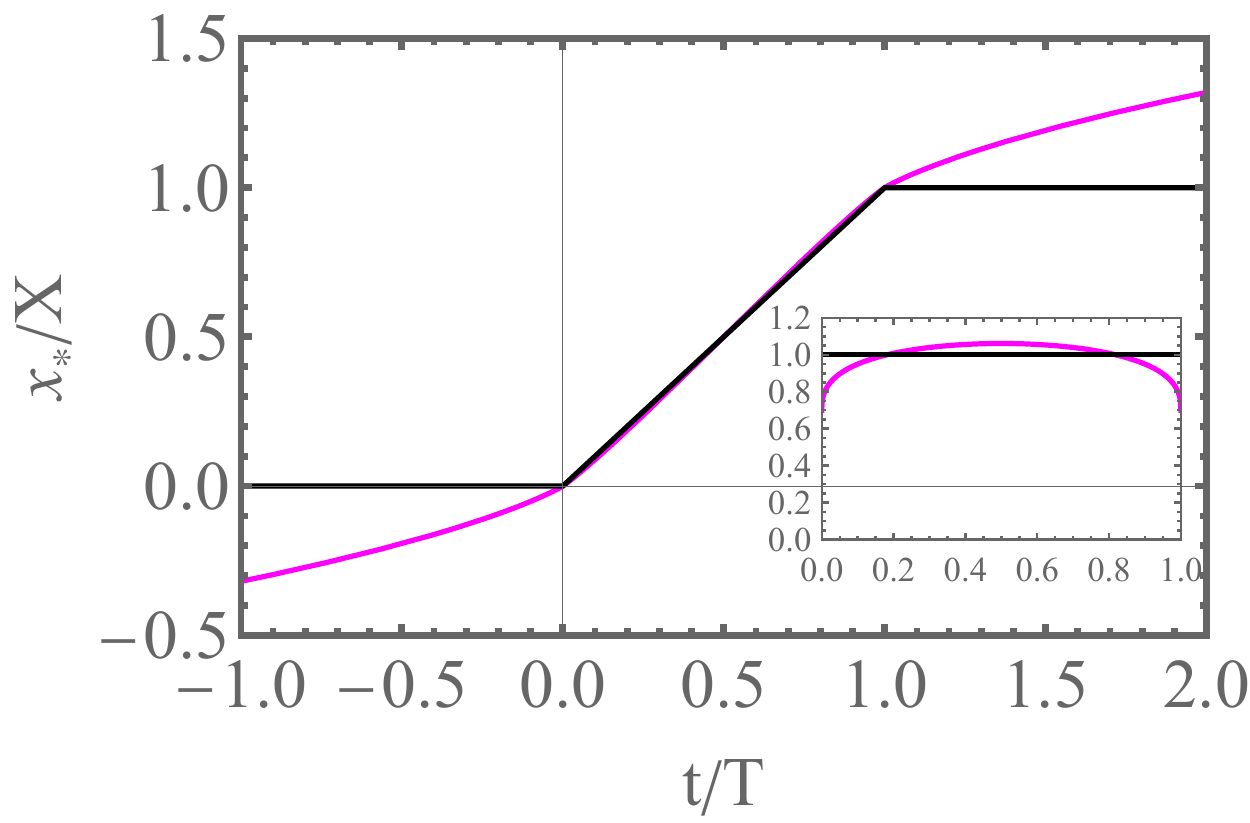}
\caption{Optimal paths $x_*(t)$ dominating the propagator of the fBm for $H=3/20$ (the blue line) and $H=7/10$ (the magenta line). Shown is $x_*(t)/X$ vs. $t/T$. For comparison, the black lines show $x_*(t)/X$ for the standard Brownian motion, $H=1/2$. The inset shows the time derivatives $dx_*/dt$ (also rescaled by $X$) vs. $t/T$ for $H=7/10$ and $H=1/2$.}
\label{xfBM}
\end{figure}

For $H<1/2$ the optimal path develops cusps at the points $t=0$ and $t=T$. For $H=1/2$ (the standard Brownian motion), $x_*(t)$ is ballistic for $0<t<T$ and takes constant values $0$ and $X$ for $t<0$ and $t>T$, respectively. When $H=1$ the optimal path becomes ballistic, $x_*(t)=(X/T)\, t$, for all $|t|<\infty$. Figure \ref{xfBM} shows examples of the optimal path for $H=3/20$ and $7/10$.

With $x_*(t)$ from Eq. (\ref{xXT}), we can now evaluate the probability distribution (\ref{Pgeneral}). Equation~(\ref{actiongeneral}) yields
\begin{equation}\label{actionresult}
S= \frac{X^2}{8 D^2 T^{4H}}\int_{-\infty}^{\infty} dt \int_{-\infty}^{\infty} dt' K(t,t') \kappa(t,T) \kappa(t',T)\,.
\end{equation}
The integral over $t'$ yields the delta-function $\delta(t-T)$ by virtue of Eq.~(\ref{inverse}). The integral over $x$ is then trivially calculated, and we obtain
\begin{equation}\label{lnP}
-\ln \mathcal{P}(X,t) \simeq S = \frac{X^2}{4 D T^{2H}}\,,
\end{equation}
as indeed to be expected from the exact Eq. (\ref{GreenfBM}).

The same short-time asymptotic (\ref{lnP}) is obtained for the probability distribution of the \emph{first passage} from $x=0$ to $x=X$. For $H=1/2$ this quantity is
known exactly due to L\'{e}vy and Smirnov, see \textit{e.g.} Ref. \cite{Redner,metzler}:
\begin{equation}\label{fp05}
\mathcal{P}(X,t) = \frac{X}{\sqrt{4 \pi D T^3}} e^{-\frac{X^2}{4 DT}}\,.
\end{equation}
For fBm the exact first-passage distribution is unknown. To our knowledge, even the short-time asymptotic (\ref{lnP}) has not been found previously.

\subsection{Three-point distribution}
\label{three}

We now use the geometrical optics to evaluate the three-point probability distribution  of a trajectory which starts at the origin at $t=0$ and is constrained to reach a position $x=X_1$ at $t=T_1$ and a position $x=X_2$ at $t=T_2$ (with unordered $T_1$  and $T_2$ which we suppose for simplicity to be both positive). To this end  we have to introduce an additional Lagrange multiplier, which changes the right-hand-side of Eq. (\ref{B100}) to
\begin{equation}
\!\int_{-\infty}^{\infty} dt^\prime\,K(t,t^\prime)x(t^\prime) \!=
\!\frac{\lambda}{2} \delta(t-T_1) + \!\frac{\mu}{2} \delta(t-T_2) \,.
\end{equation}
Multiplying the both sides of this equation by $\kappa(t,\tau)$ and  integrating
over $t$ using Eq. (\ref{inverse}), we determine the functional form of the optimal path:
\begin{equation}
\label{3opt}
x_*(t) = \frac{\lambda}{2} \kappa(T_1,t) + \frac{\mu}{2} \kappa(T_2,t) \,.
\end{equation}
Recalling that $x_*(T_1) = X_1$ and $x_*(T_2) = X_2$, we find $\lambda$ and $\mu$:
\begin{align}
\lambda & = \frac{2 \left(\kappa(T_1,T_2) X_2 - \kappa(T_2,T_2) X_1\right)}{\kappa^2(T_1,T_2) - \kappa(T_1,T_1) \kappa(T_2,T_2)}  \,, \label{lambda3}\\
\mu & = \frac{2 \left(\kappa(T_1,T_2) X_1 - \kappa(T_1,T_1) X_2\right)}{\kappa^2(T_1,T_2) - \kappa(T_1,T_1)  \kappa(T_2,T_2)} \,,
\label{mu3}
\end{align}
which completes the calculation of $x_*(t)$.

Now we insert Eqs. (\ref{3opt})-(\ref{mu3}) into Eq. (\ref{actiongeneral}). Splitting the double integral into four integrals and making use of the identity (\ref{inverse}), we arrive at
\begin{equation}
S = \frac{\lambda}{4} X_1 + \frac{\mu}{4} X_2 \,.
\end{equation}
After some algebra we obtain
\begin{eqnarray}
  &-&\!\!\ln \mathcal{P} \!\simeq S \!=\! \frac{1}{4 D(1 - g^2)} \left(\frac{X_1^2}{T_1^{2H}} +
\frac{X_2^2}{T_2^{2 H}} - \frac{2 g X_1 X_2}{D T_1^H T_2^H}\right), \nonumber\\
 &&\text{where}\quad g = \frac{T_1^{2H} + T_2^{2H} - |T_1 - T_2|^{2H}}{2 T_1^H T_2^H} \,.
  \label{3point}
\end{eqnarray}
This expression, up to a proper normalization, coincides with the known
exact result (see, \textit{e.g.} Ref. \cite{alessio}). $N$-point probability distributions for $N>3$ can
be calculated in a similar way, by introducing additional delta-functions with Lagrange multipliers.

\subsection{Distribution of the maximum of fractional Brownian bridge and fractional Brownian excursion}
\label{distmax}

Extreme-value statistics is an important area of probability theory, with applications ranging from physics and engineering to climate, finance and sports \cite{Fortin,MPS}. Here we use the findings of Sec. \ref{three} to determine the large-deviation statistics
of the maximum value $M$ of fractional Brownian bridge (fBb), and of fractional Brownian excursion (fBe), on the time interval $0<t<T$. Here is how these processes are defined.  Apart from the condition $x(0)=0$, which is obeyed automatically, each of these two processes obeys the condition $x(t=T)=0$. The difference between the two is in that the fBe is also required to stay positive on the interval $0<t<T$, while the fBb is not.

\begin{figure}[ht]
\includegraphics[width=0.3\textwidth,clip=]{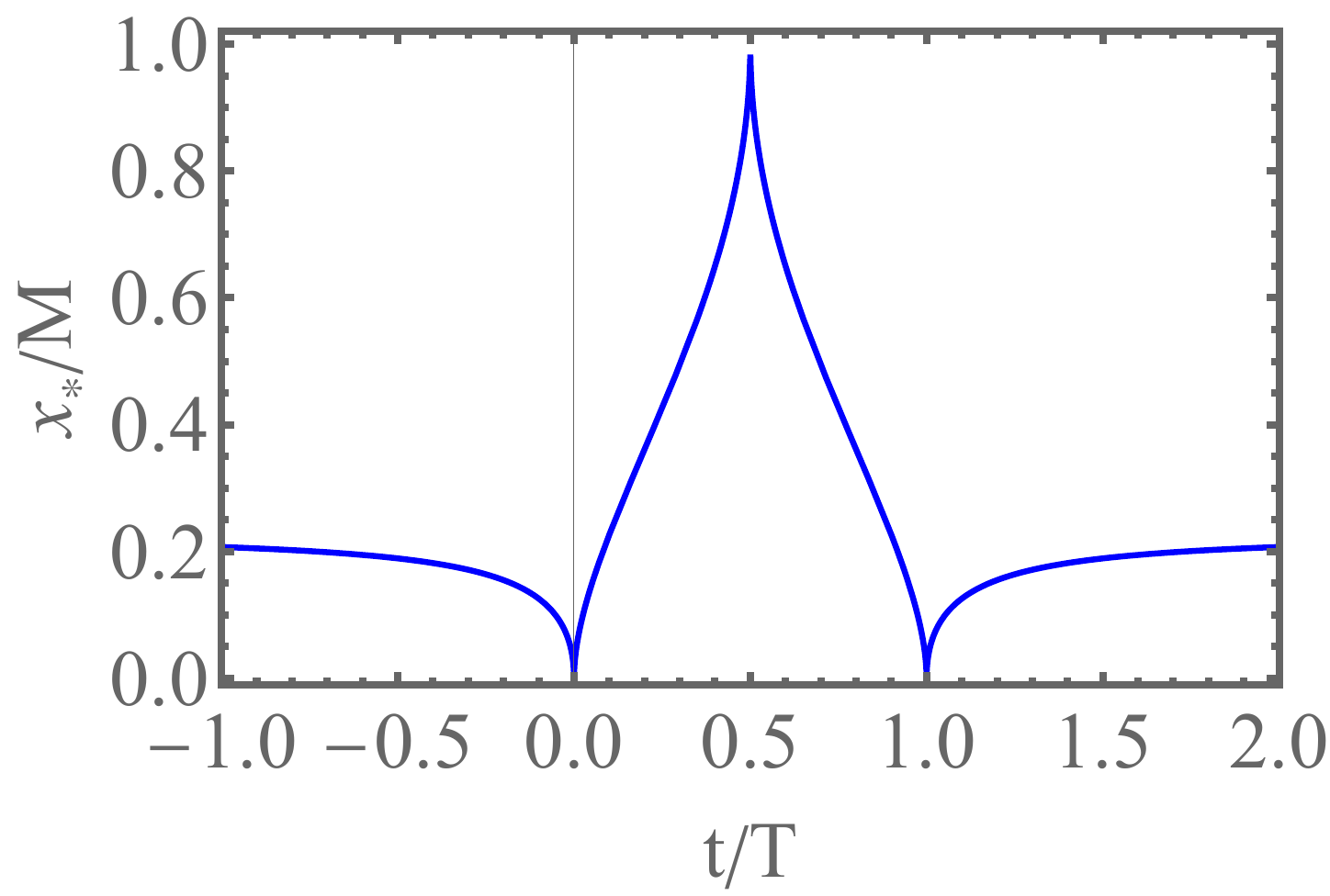}
\includegraphics[width=0.3\textwidth,clip=]{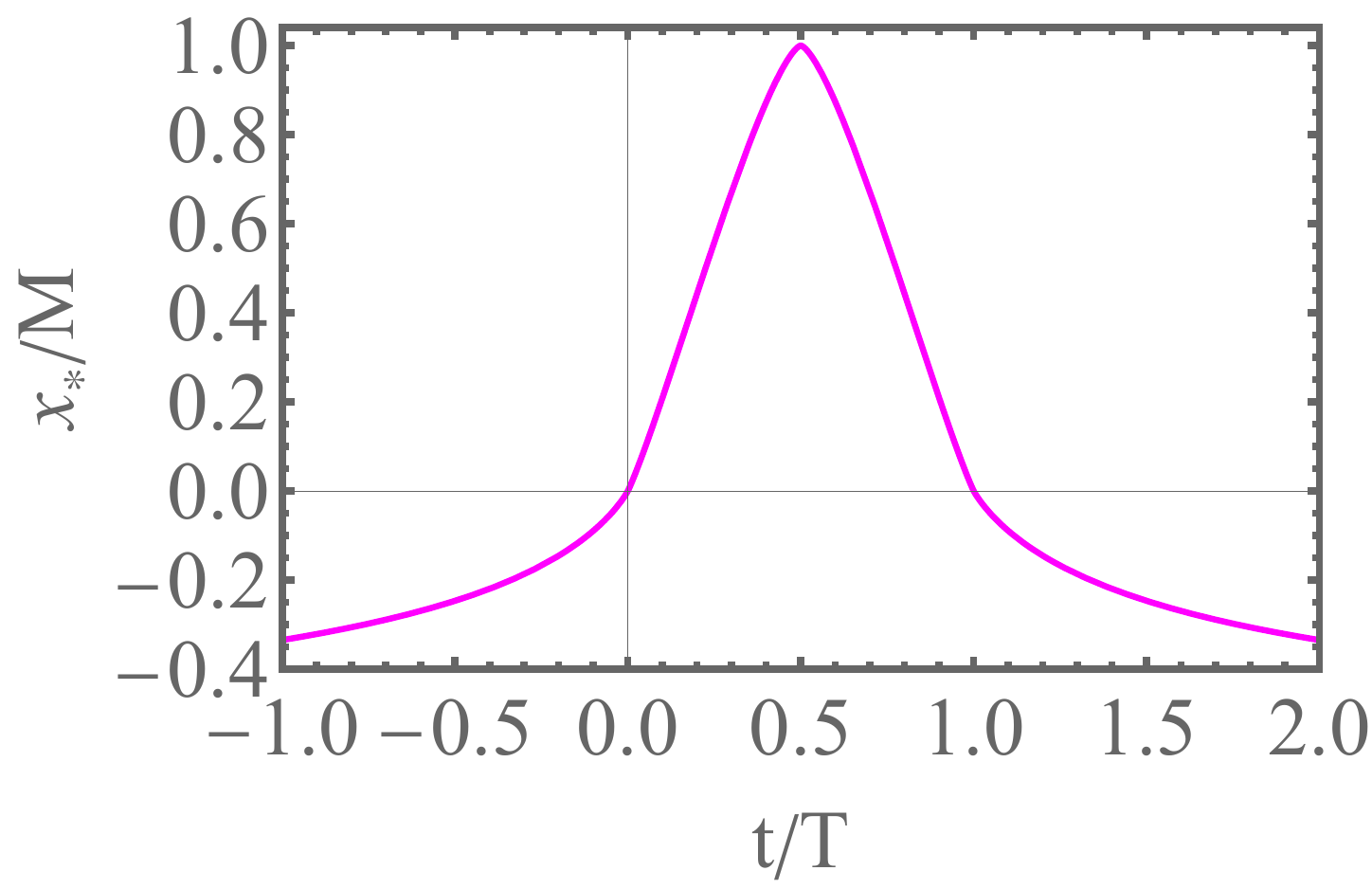}
\includegraphics[width=0.3\textwidth,clip=]{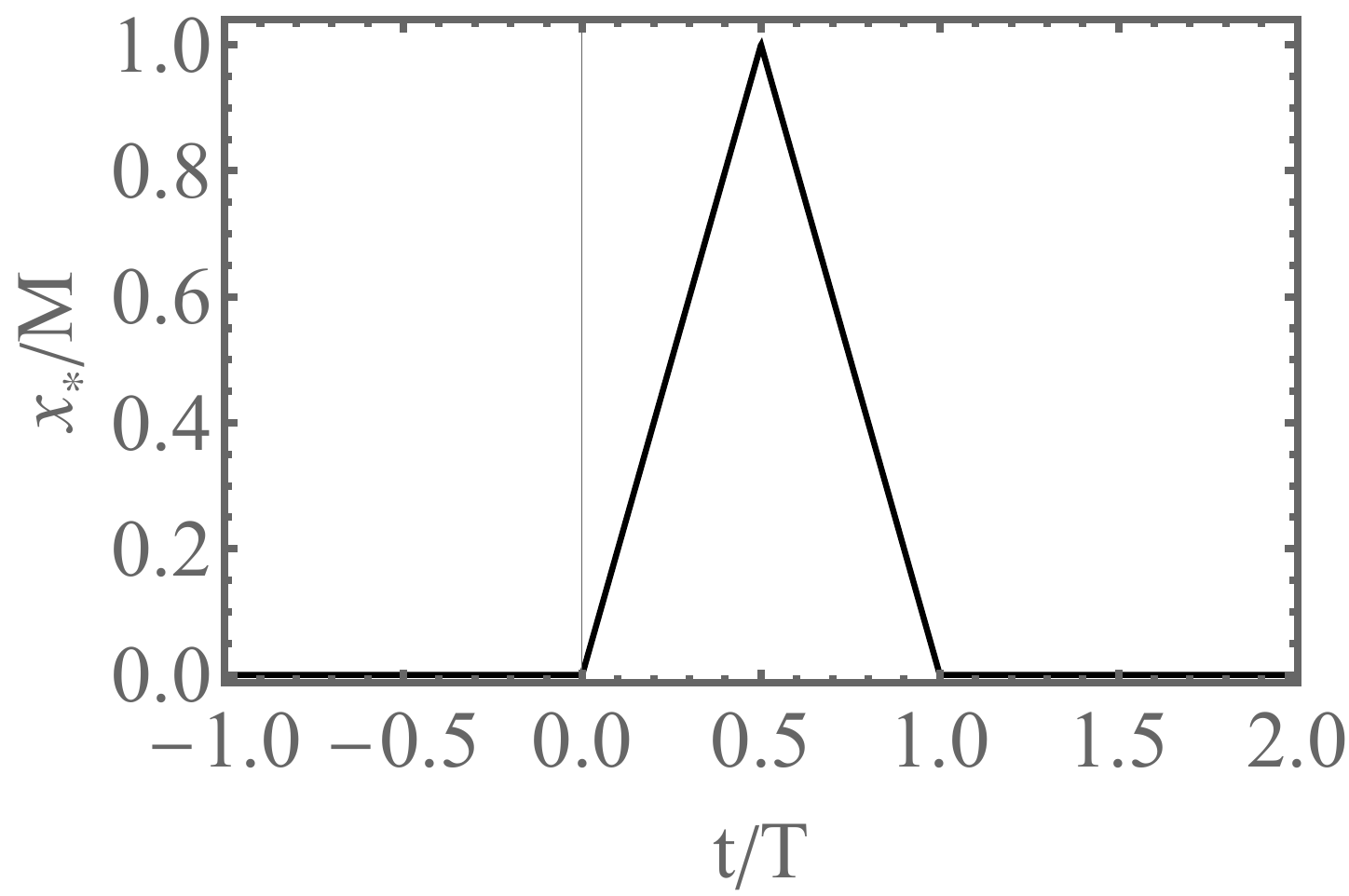}
\caption{Optimal paths dominating fBb and fBe conditioned on reaching a large maximum $M$ on the interval $0<t<T$, see Eq.~(\ref{areadef}). Shown is $x_*(t)/M$ versus $t/T$ for $H=1/4$ (the blue line) and $H=3/4$ (the magenta line). For comparison, the black line shows the optimal path for the standard Brownian motion, $H=1/2$.}
\label{max}
\end{figure}

The geometrical optics applies when $M$ is large or $T$ is small. To determine the optimal path $x_*(t)$ from the results of Sec. \ref{three}, we can set $X_1=M$, $X_2=0$, $T_2=T$ and \emph{minimize} the action $S$ in Eq.~(\ref{3point}) with respect to the \textit{a priori} unknown value of the time $T_2$ of the maximum, $x(T_2) = M>0$, where $0<T_2<T$. This minimization gives $T_2=T/2$.
The resulting minimum action yields the previously unknown large deviation tail
\begin{equation}\label{maximum}
-\ln \mathcal{P}(M,T) \simeq S = \frac{M^2}{\left(4^{1-H}-1\right) D T^{2H}}\,.
\end{equation}
The optimal path $x_{*}(t)$ is described by Eq.~(\ref{3opt}) with
\begin{equation}\label{lmM}
\lambda = \frac{4 M}{\left(4^{1-H}-1\right) D T^{2 H}}\,,\quad \mu = -\frac{2 M }{\left(4^{1-H}-1\right) D T^{2 H}}\,.
\end{equation}
Figure \ref{max} shows the optimal path $x_*(t)$ for $H=1/4$, $3/4$ and $1/2$. Noticeable is the cusp at $t=T/2$ for $H<1/2$.

Since the optimal path does not cross $x=0$ for $0<t<T$, it is shared by both fBb and fBe. Correspondingly, the large-$M$ tail (\ref{maximum}) of the distribution is the same for both processes up to pre-exponential factors which are left beyond the leading-order geometrical optics.

As one can see from Eq.~(\ref{maximum}), the $H$-dependence of $\mathcal{P}$ is quite strong. $\mathcal{P}$ is finite in the limit of $H\to 0$, but it vanishes in the limit of $H \to 1$, when $\lambda$ and $\mu$ in Eq.~(\ref{lmM}) diverge, and the optimal path $x_*(t)$ becomes singular. The latter fact is to be expected on the physical grounds. Indeed, for $H>1/2$ the correlations of the increments of the fBm are positive. As $H$ approaches $1$ the correlations become so strong that it is impossible to ``bend" the trajectory and make it return to $x=0$ at $t=T$.

For the normal fBm ($H=1/2$) Eq.~(\ref{maximum}) yields
\begin{equation}\label{maximum05}
-\ln \mathcal{P}(M,T) \simeq \frac{M^2}{D T}\,.
\end{equation}
which agrees, up to a pre-exponent, with the known exact results for the Brownian bridge and Brownian excursion, see \textit{e.g.} Ref. \cite{SLD}.

The tail~(\ref{maximum}) of the bridge maximum distribution was previously obtained by Delorme and Wiese \cite{Delorme} who considered \emph{one-sided} fBm. The authors arrived at this result via a minimization procedure that they called an ``heuristic argument". In fact, what they did was very close in spirit to geometrical optics. 
That the tails for the one-sided and two-sided fBm  coincide in this case is just one example of a general property, as we discuss in section \ref{discussion}.

\section{Area under fBm}
\label{areas}

In this section we consider the area distribution
under fBm on a finite time interval $0<t<T$,
\begin{equation}\label{areadef}
A_T \equiv \int_0^T x(t) \,dt = A\,,
\end{equation}
in three different settings. We will start from the simplest case, where Eq.~(\ref{areadef}) is the \emph{only} constraint, and the area  distribution is known. This will be an additional test of the geometrical optics.

\subsection{No constraints except Eq.~(\ref{areadef})}
\label{noconstraints}

Since the relation (\ref{areadef}) is linear,  the distribution $\mathcal{P}(A,T)$ that we are after is Gaussian, and it can be easily calculated exactly. Indeed, it has zero mean, and the variance
\begin{equation}\label{varTexact}
\langle A_T^2\rangle = \int_0^T \int_0^T dt \,dt' \kappa(t,t') = \frac{D T^{2+2H}}{1+H}\,.
\end{equation}
Therefore,
\begin{equation}\label{distnoconstraint}
\mathcal{P}(A,T)=\sqrt{\frac{1+H}{2\pi D T^{2+2H}}} \exp\left[-\frac{(1+H) A^2}{2 D T^{2+2H}}\right].
\end{equation}

\begin{figure}[ht]
\includegraphics[width=0.3\textwidth,clip=]{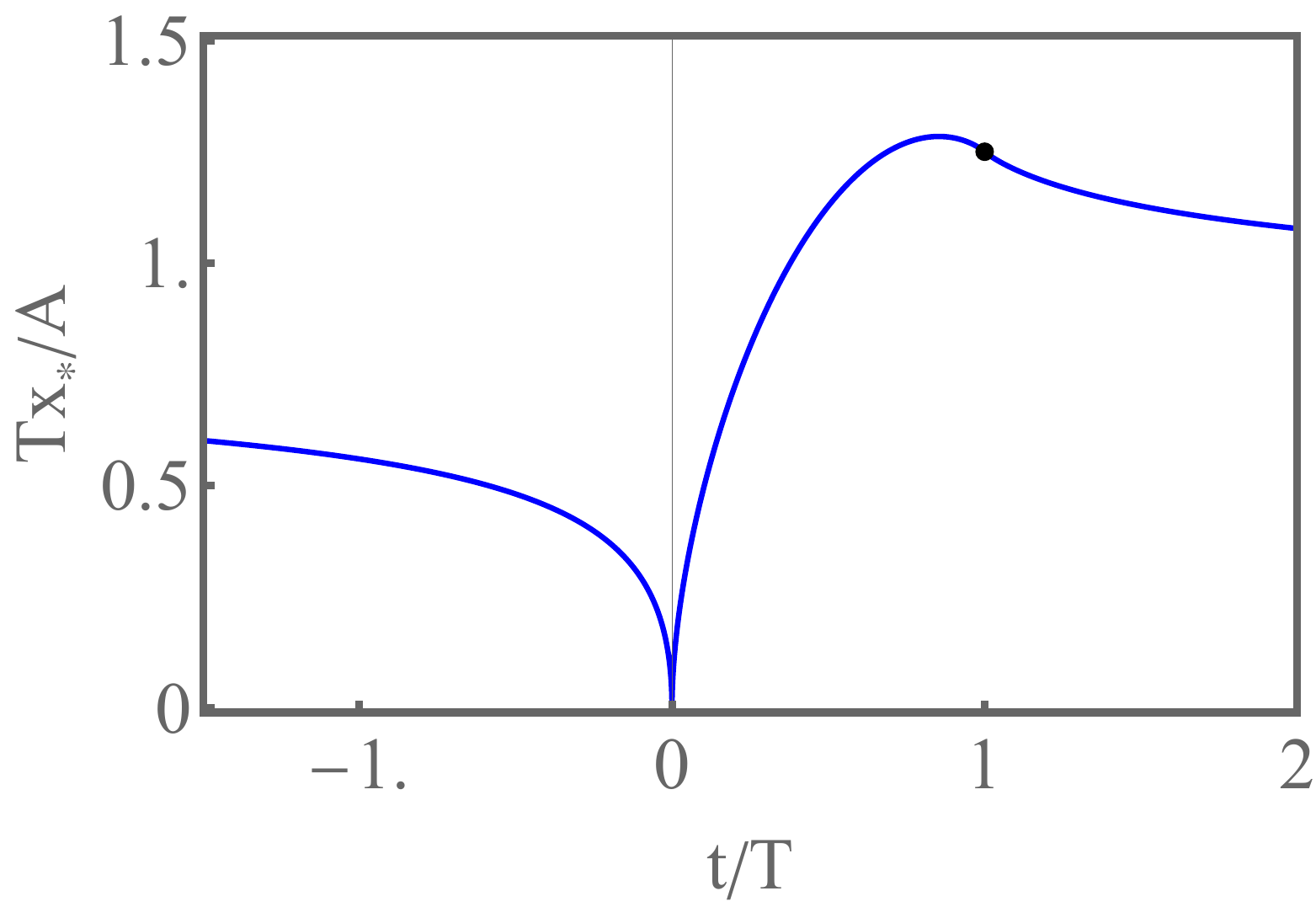}
\includegraphics[width=0.3\textwidth,clip=]{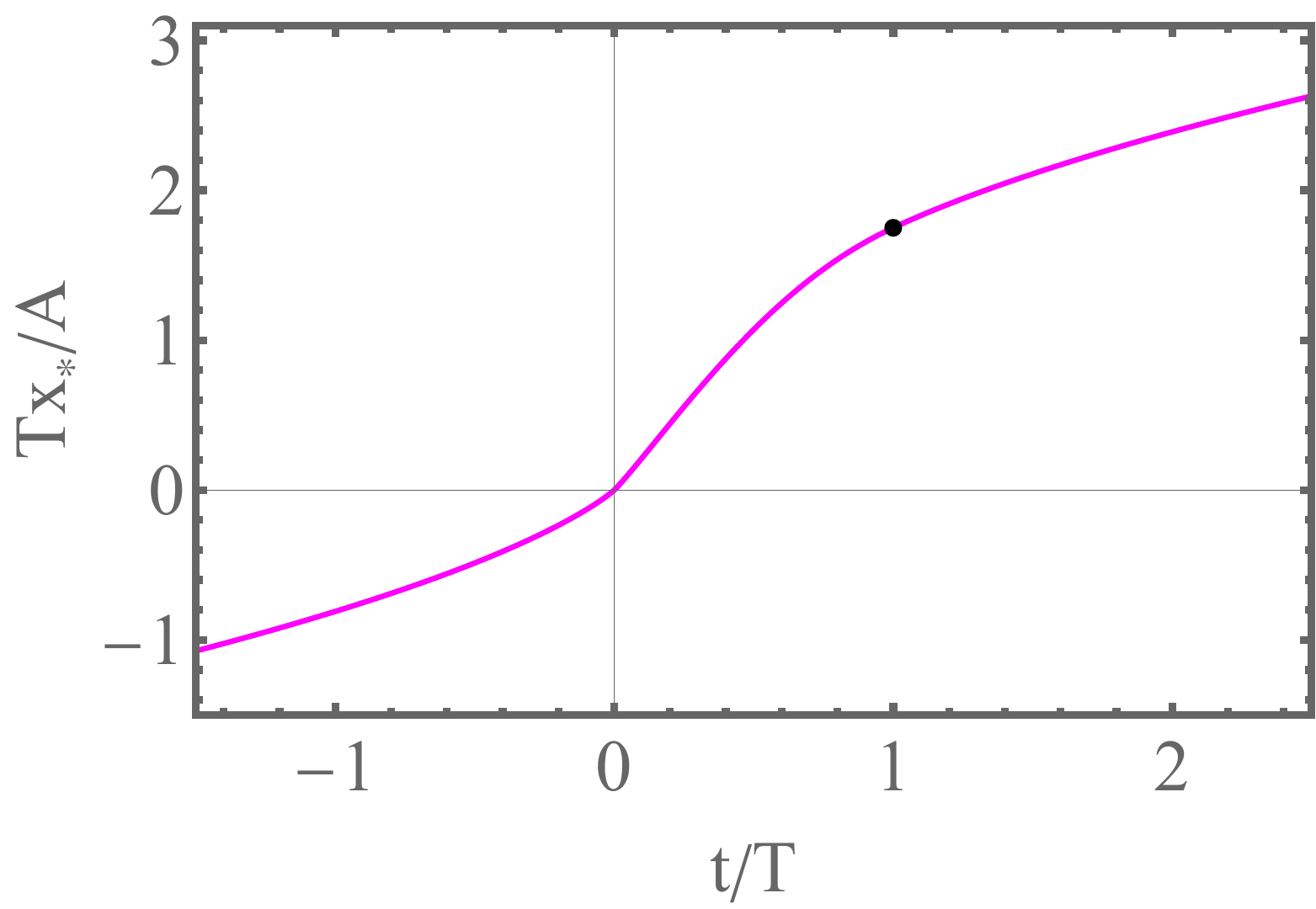}
\includegraphics[width=0.3\textwidth,clip=]{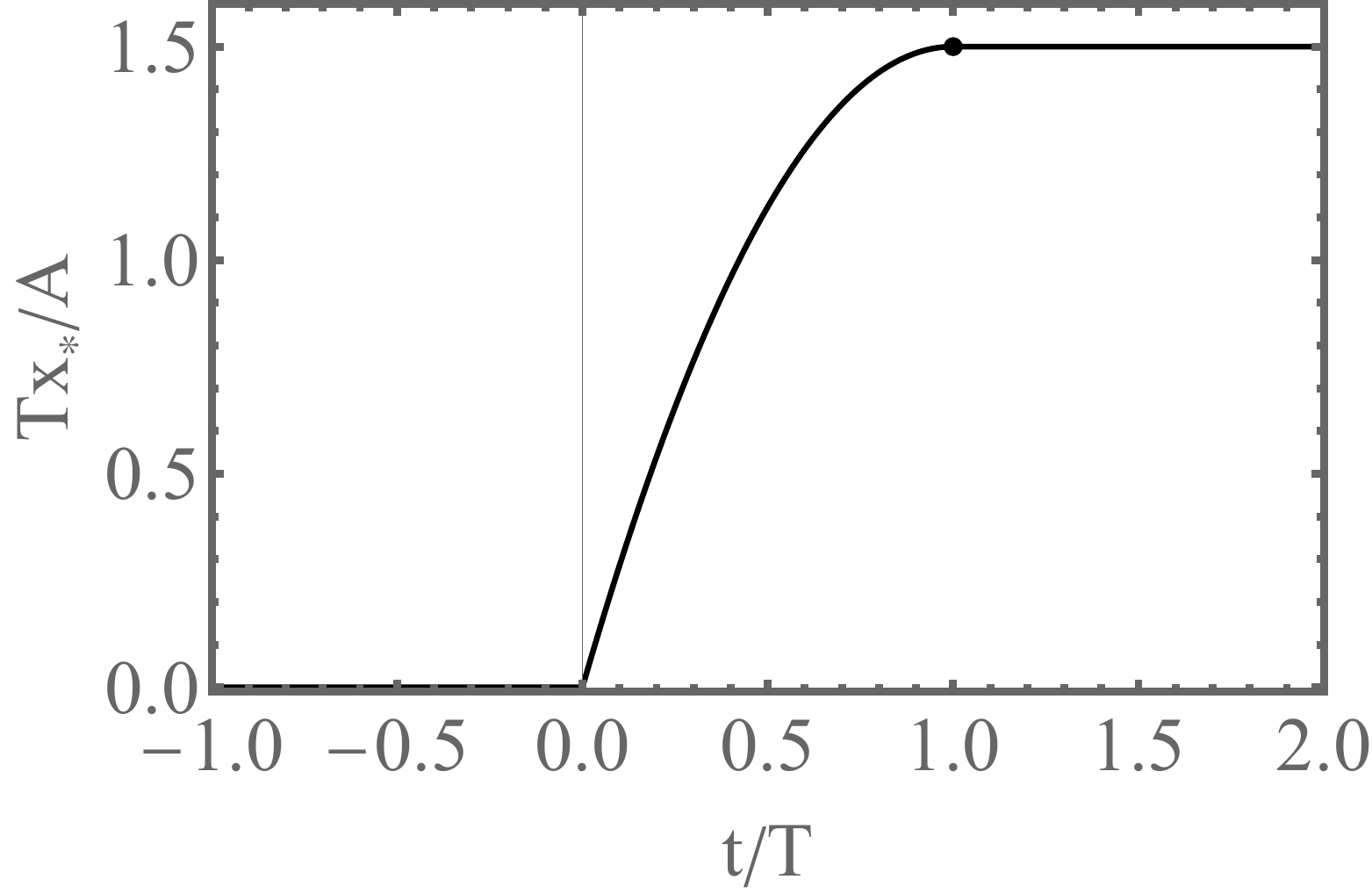}
\caption{Optimal paths $x_*(t)$ dominating the fBm conditioned on a given area $A$ on a specified interval $0<t<T$, see Eq.~(\ref{areadef}). Shown is $T x_*(t)/A$ versus $t/T$ for $H=1/4$ (the blue line) and $H=3/4$ (the magenta line). For comparison, the black line shows the optimal path for the standard Brownian motion,
$H=1/2$.  The dots indicate the optimal values of $x(T)$.}
\label{noconstraintfig}
\end{figure}

Now we rederive this result by assuming a
very large $A$ or short times $T$ and using geometrical optics. The constrained action to be minimized
is
\begin{eqnarray}
 S_{\lambda}[x(t)]&=& \frac{1}{2}\int_{-\infty}^{\infty} dt \left[\int_{-\infty}^{\infty} dt^\prime  K(t,t^\prime) x(t) x(t^\prime) \right. \nonumber\\
  &-& 
  \mu h_T(t)  x(t) \bigg]\,,
\label{functional2a}
\end{eqnarray}
where $h_T(t)$ is the rectangle function: $h_T(t)=1$ for $0<t<T$ and zero otherwise, and $\mu$ is the Lagrange multiplier to be ultimately expressed through $A$ and $T$. The minimization leads to the linear equation
\begin{equation}\label{B200a}
\!\int_{-\infty}^{\infty} dt^\prime\,K(t,t^\prime)x(t^\prime) = \frac{\mu}{2} h_T(t).
\end{equation}
Multiplying the both sides of this equation by $\kappa(t,\tau)$, integrating over $t$ and using the identity  (\ref{inverse}),  we obtain
\begin{equation}\label{x200a}
x_*(t) = \frac{\mu}{2}  \int_0^T \kappa(t,\tau) d\tau \,.
\end{equation}
Now the optimal path is given by an integral of the covariance.  Demanding that the area (\ref{areadef}) be equal to $A$, we determine the Lagrange multiplier:
\begin{equation}\label{mu}
\mu =\frac{2 A (1+H)}{D T^{2+2 H}}\,.
\end{equation}
The integral in Eq.~(\ref{x200a}) can be evaluated explicitly, but the resulting expressions are  too cumbersome to be presented here. Three examples of the optimal path $x_*(t)$ are depicted in Fig. \ref{noconstraintfig}.

For the standard Brownian motion $H=1/2$ the optimal path is
\begin{numcases}
{{x_*(t)} =} 0,\!\! & $t <0$, \nonumber \\
\frac{3 A t (2 T-t)}{2 T^3},\!\!& $0<t<T$, \label{BMun}\\
\frac{3A}{2T},\!\!& $t>T$. \nonumber\\
   \nonumber
\end{numcases}
For $H=1$ the optimal path is again ballistic, $x_*(t)=2 A t/T^2$, for all $|t|<\infty$.

Using the optimal path, described by Eqs.~(\ref{x200a}) and (\ref{mu}), we can evaluate the probability distribution, Eqs.~(\ref{actiongeneral}) and (\ref{Pgeneral}). In this way we arrive at a quadruple integral. Again, the identity (\ref{inverse}) reduces one integration. The resulting delta-function reduces one more integration, and we arrive at
\begin{equation}\label{lnP1a}
 -\ln \mathcal{P}(A,t) \simeq  \frac{\mu^2}{8} \int_{0}^{T} dt \int_0^T dt'\, \kappa(t,t') \,,
\end{equation}
with $\mu$ from Eq.~(\ref{mu}). Evaluating this double integral, we obtain
\begin{equation}\label{lnP1b}
 -\ln \mathcal{P}(A,t) \simeq  \frac{(1+H)A^2}{2 D  T^{2 +2H}}
\end{equation}
in agreement with the exact result (\ref{distnoconstraint}).

\vspace{0.5cm}

\subsection{Area under fractional Brownian bridge and fractional Brownian excursion}
\label{fBBfBE}

Now we return to the fBb and fBe and study the distribution $\mathcal{P}(A,T)$ of the area (\ref{areadef}) under them.
For $H=1/2$ these two area distributions are known exactly. The area distribution under the standard Brownian excursion ($H=1/2$) is known as the Airy distribution. Since its discovery almost four decades ago \cite{Darling,Louchard}, the
Airy distribution  has appeared in multiple problems
in physics and computer science, see Ref. \cite{MajumdarComtet} for a review and Ref. \cite{Agranovetal} for recent developments.

Area distributions under fBb and fBe are unknown, and finding them is a natural next step of theory. Here we suppose that the specified area $A$ is very large (or $T$ is very small) which enables us to use geometrical optics. Now, in addition to the constraint on the area, Eq.~(\ref{areadef}), we must also impose the constraint $x(T)=0$. We impose the latter constraint by introducing an additional Lagrange multiplier, so the constrained action to be minimized
is
\begin{eqnarray}
 S_{\lambda}[x(t)]&=& \frac{1}{2}\int_{-\infty}^{\infty} dt \left[\int_{-\infty}^{\infty} dt^\prime  K(t,t^\prime) x(t) x(t^\prime) \right. \nonumber\\
  &-&  \mu h_T(t)  x(t)- \lambda x(t) \delta(t-T)\bigg]\,,
\label{functional2}
\end{eqnarray}
where $\mu$ and $\lambda$ are two Lagrange multipliers to be ultimately expressed through $A$ and $T$. The minimization leads to the linear equation
\begin{equation}\label{B200}
\!\int_{-\infty}^{\infty} dt^\prime\,K(t,t^\prime)x(t^\prime) = \frac{\mu}{2} h_T(t)+\frac{\lambda}{2} \delta(t-T)\,.
\end{equation}
Multiplying both parts of this equation by $\kappa(t,\tau)$, integrating over $t$ and swapping  $\tau$ and $t$, we obtain
\begin{equation}\label{x200}
x_*(t) = \frac{\mu}{2}  \int_0^T \kappa(t,\tau) d\tau + \frac{\lambda}{2} \kappa(t,T)\,.
\end{equation}
Here the optimal path is a linear combination of the covariance and its integral over time.

\begin{figure}[ht]
\includegraphics[width=0.3\textwidth,clip=]{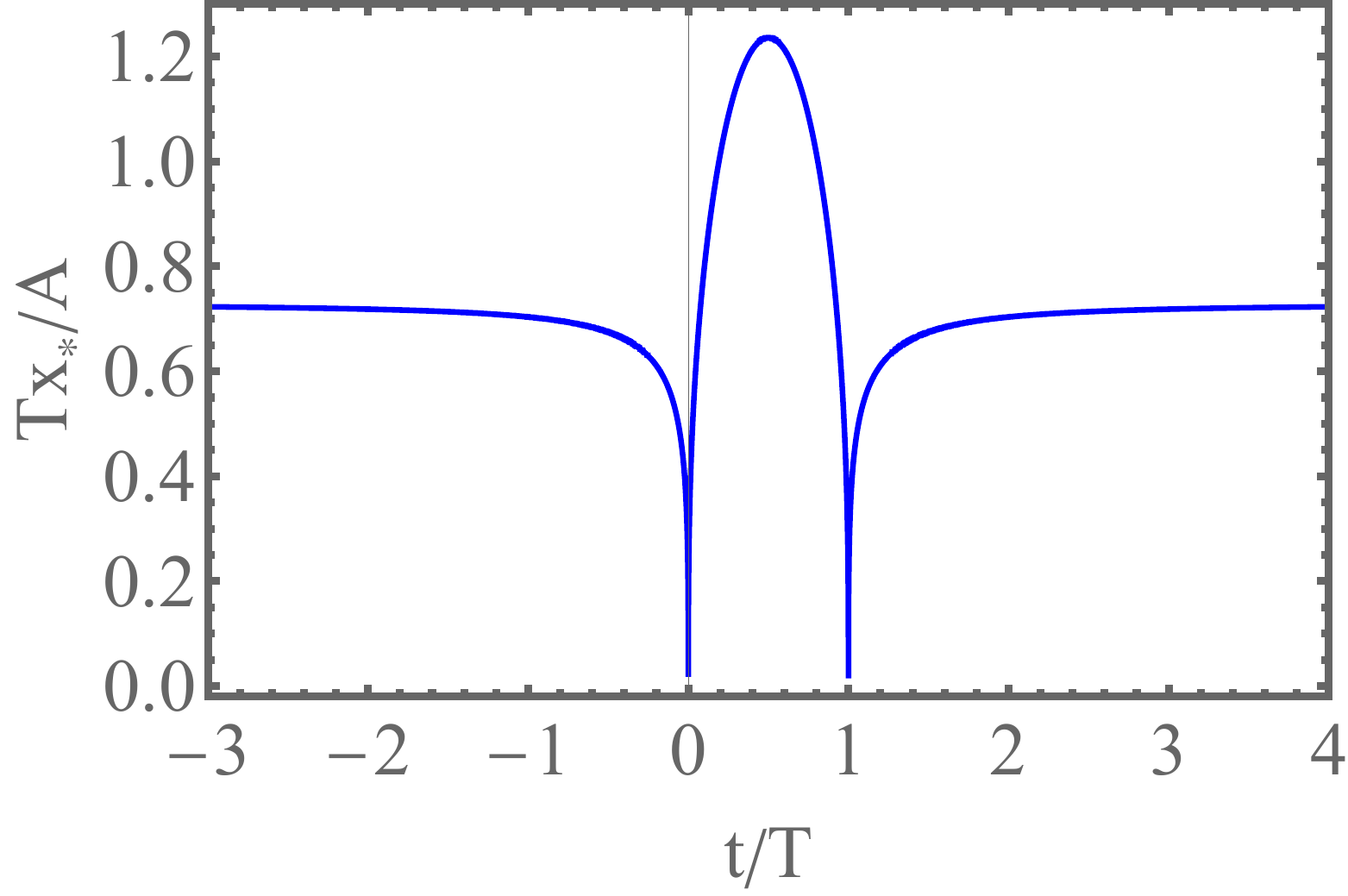}
\includegraphics[width=0.3\textwidth,clip=]{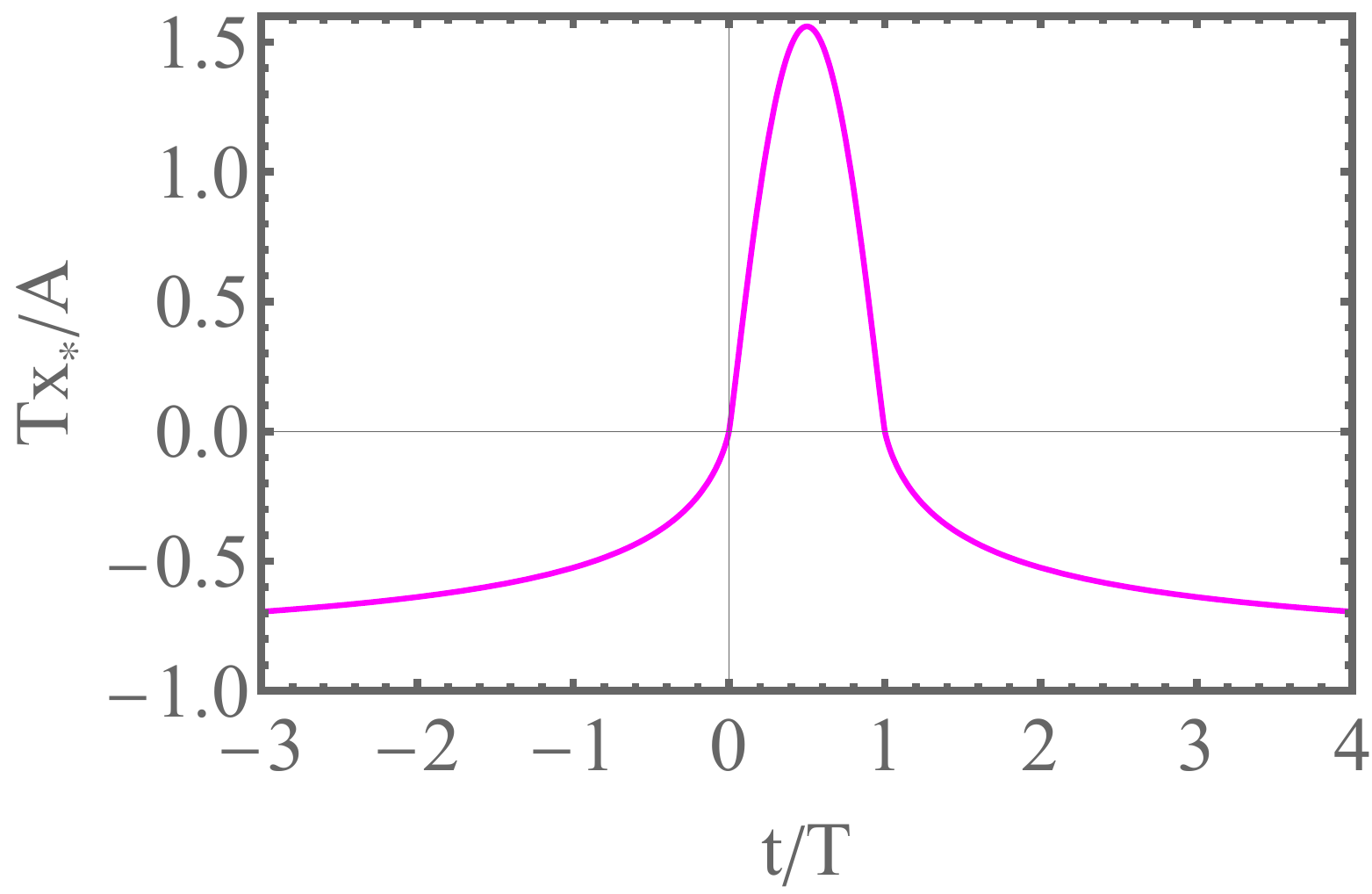}
\includegraphics[width=0.3\textwidth,clip=]{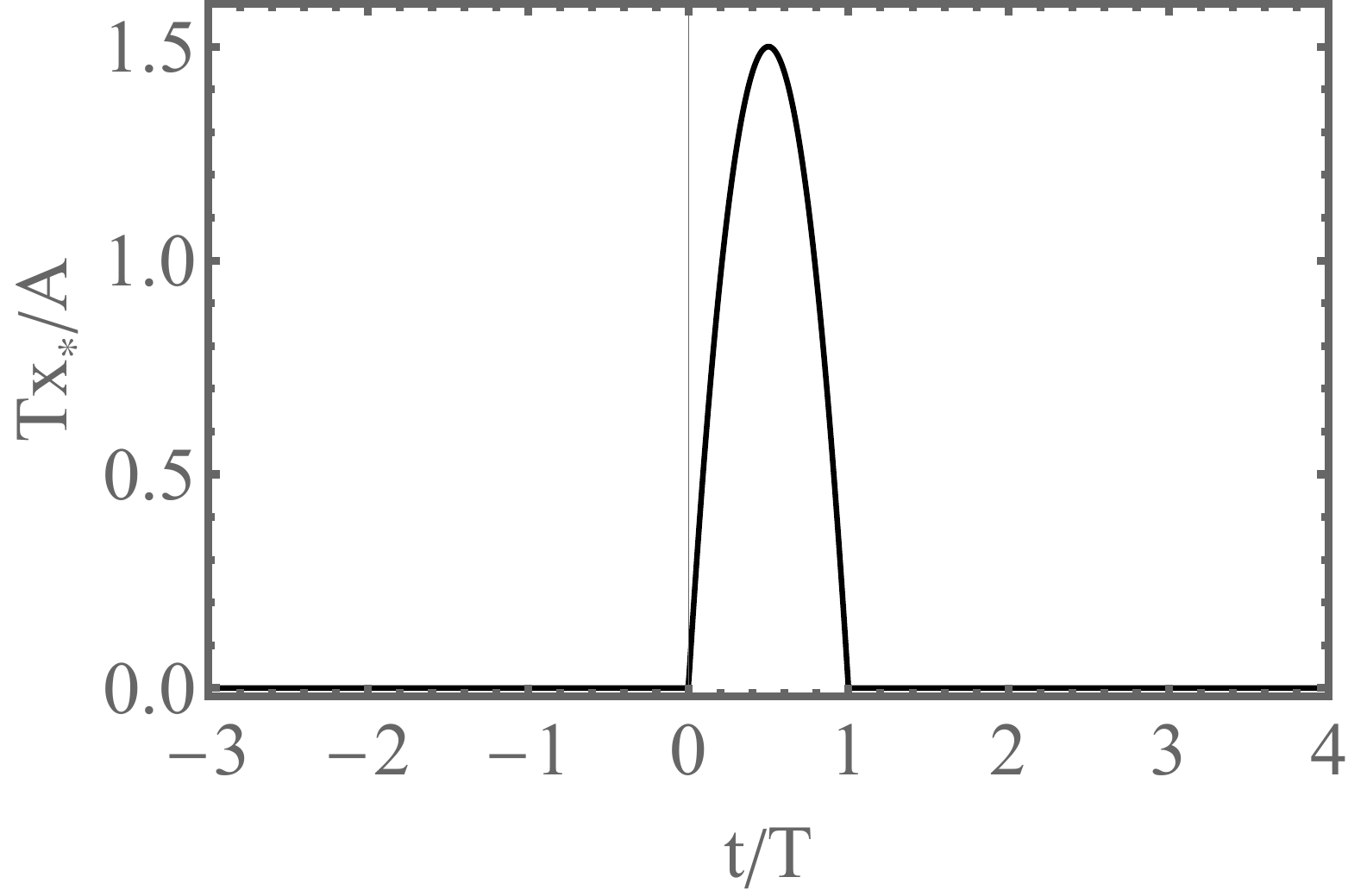}
\caption{(Rescaled) optimal paths dominating the fBe conditioned on a given area $A$ for $H=3/20$ (the blue line) and $H=7/10$ (the magenta line). For comparison, the black line shows the optimal path for the standard Brownian motion,
$H=1/2$, see Eq. \ref{BE} and Ref. \cite{Agranovetal}.}
\label{excursionfig}
\end{figure}

Now we demand that  $x_*(T)=0$ and that the area (\ref{areadef}) be equal to $A$. These two conditions are required for both the fBb and fBe, and they give
\begin{equation}\label{muandlambda}
\mu=\frac{4 A (1+H) T^{-2 H-2}}{D (1-H)} ,\quad
\lambda = -\frac{2 A (1+H) T^{-2 H-1}}{D (1-H)}\,.
\end{equation}
Three examples of the resulting optimal path (\ref{x200})  are shown in Fig. \ref{excursionfig}.

As one can see, the optimal path $x_*(t)$ stays positive on the interval $0<t<T$ for all $0<H<1$. As a result, this optimal path is shared by the  fBb and fBe, hence these two processes have the same  large-$A$ or small-$T$ asymptote of $\mathcal{P}(A,t)$ up to pre-exponential factors.  We will determine this asymptote shortly.

For standard Brownian excursion, $H=1/2$, the optimal path (\ref{x200}) becomes quite simple,
\begin{numcases}
{{x_*(t)} =} 0,\!\! & $t <0$, \nonumber \\
\frac{6A t(T-t)}{T^3},\!\!& $0<t<T$, \label{BE}\\
 0,\!\!& $t>T$, \nonumber\\
   \nonumber
\end{numcases}
in agreement with Ref. \cite{Agranovetal}.

With  the optimal path (\ref{x200}) at hand, we can calculate the action (\ref{actiongeneral}). We again split the integral into 4 integrals
and use the definition (\ref{inverse}) of the inverse kernel in the integration over $t'$ in each of the integrals. The resulting $A\gg \sqrt{D}T^{1+H}$ tail of the area distribution is
\begin{equation}\label{probfBE}
-\ln \mathcal{P}(A,T) \simeq S=\frac{1+H}{1-H} \frac{A^2}{D T^{2+2H}}\,.
\end{equation}
For $H=1/2$ we obtain $-\ln \mathcal{P}(A,T) \simeq 3 A^2/(D T^3)$ in perfect agreement with the known results for the Brownian bridge (where the distribution is Gaussian) and the Brownian excursion, where it describes the large-$A$ tail of the Airy distribution \cite{MajumdarComtet,Agranovetal}. Similarly to the distribution of the maximum (\ref{maximum}), see Sec. \ref{distmax}, the distribution (\ref{probfBE}) is finite as $H \to 0$ and vanishes as $H \to 1$.

\subsection{First-passage area under fBm}
\label{fparea}

First-passage Brownian functionals (that is, functionals of the Brownian motion defined up to the time of its first passage to a certain point in space) have attracted much recent attention, see Ref. \cite{MM2020a} and references there. The particular case of the first-passage Brownian functionals -- the first-passage area -- has found applications ranging from queueing theory and combinatorics,
to the statistics of avalanches in self-organized criticality \cite{KM}. Here we extend this line of work to the fBm. Suppose that the fBm starts at $t=0$ at a specified distance $L>0$ from the origin:  $x(t=0)=L$. We will be interested in the distribution $\mathcal{P}(A,L)$ of the area (\ref{areadef}), where $T$ now is the first-passage time to $x=0$.
For normal diffusion, $H=1/2$, this distribution is known \cite{KM,MM2020a}:
\begin{equation}\label{PAexact}
\mathcal{P}(A,L) = \frac{L}{3^{2/3} \Gamma
   \left(1/3\right) (D A^4)^{1/3}}\, \exp\left(-\frac{L^3}{9 D A}\right)\,,
   \end{equation}
where $\Gamma(\dots)$ is the gamma function. Noticeable is the essential singularity at $A\to 0$.

For $H\neq 1/2$ the distribution is unknown, and here we will evaluate its $A\to 0$ asymptotic by using geometrical optics. It is convenient to reverse time, $t \to T-t$, where $T$ is \textit{a priori} unknown. Now the fBm starts at the origin at $t=0$ and reaches $x=L$ at $t=T$ without crossing $x=0$ for $0<t<T$.

The distribution $\mathcal{P}(A,L)$ is contributed to by paths with different first-passage times $T$. Within the geometrical optics approach, our search for the optimal path must now involve an additional optimization: finding the \textit{a priori} unknown optimal value of $T$ which minimizes the action (\ref{actiongeneral}).

\begin{figure}[ht]
\includegraphics[width=0.3\textwidth,clip=]{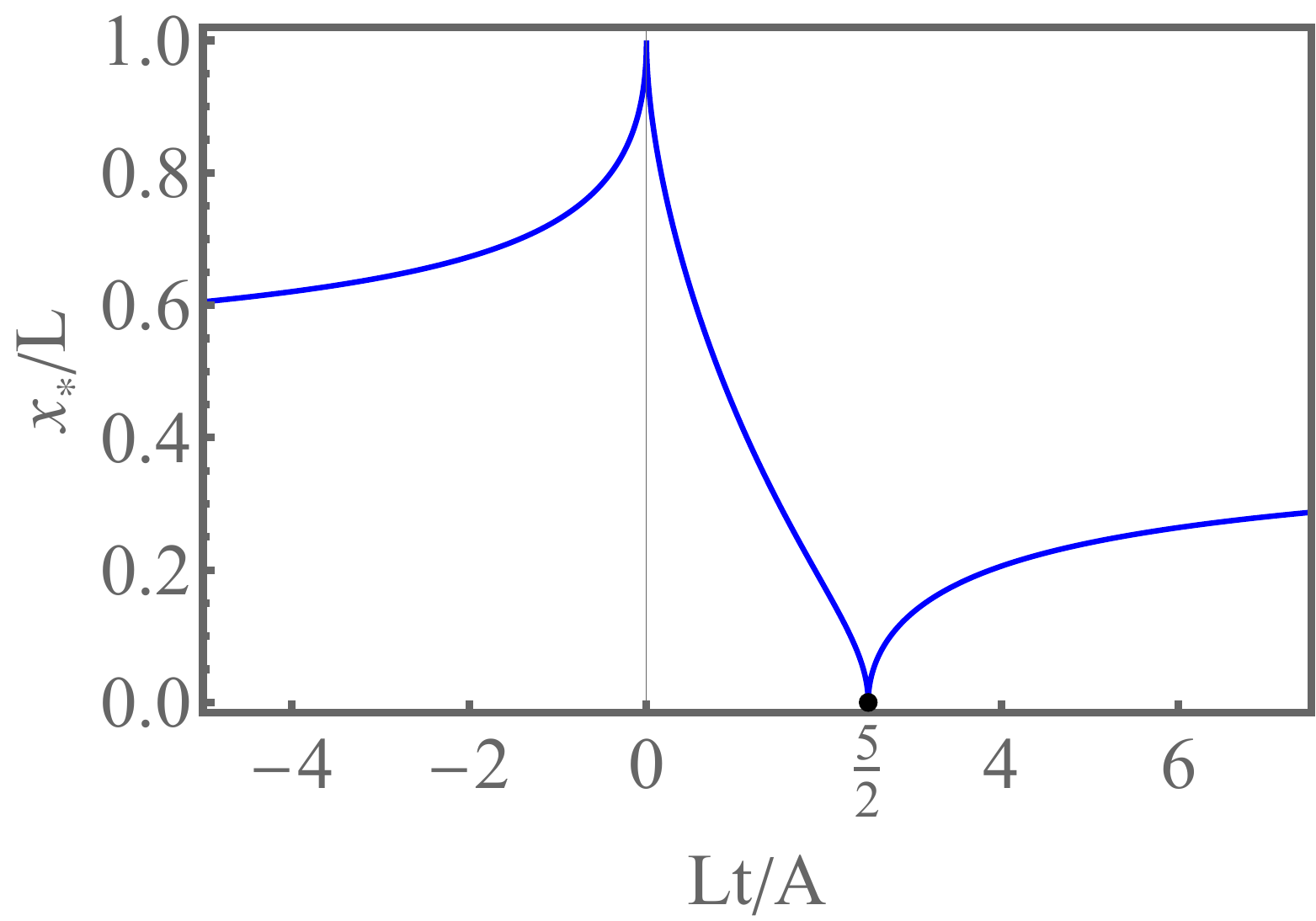}
\includegraphics[width=0.3\textwidth,clip=]{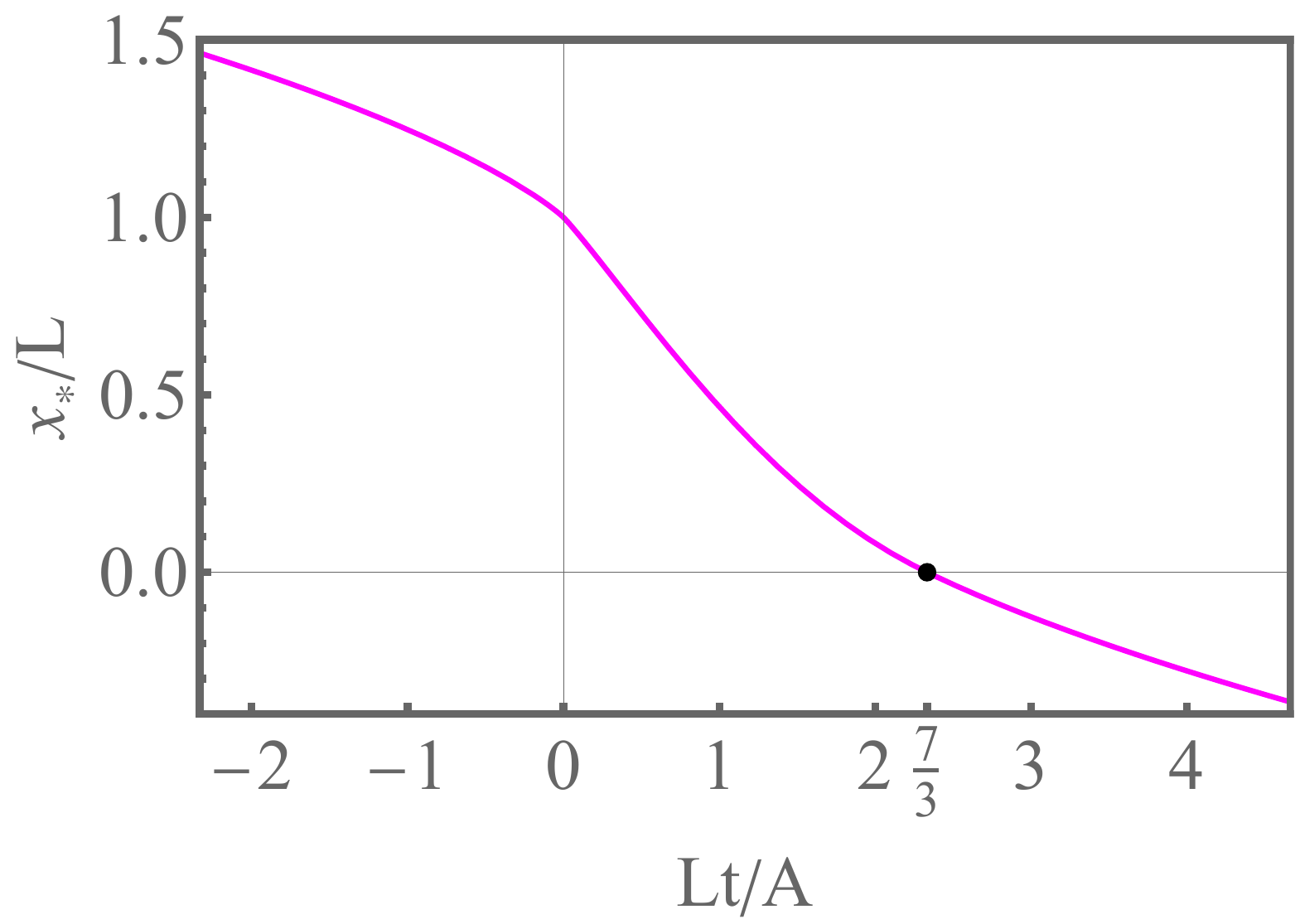}
\includegraphics[width=0.3\textwidth,clip=]{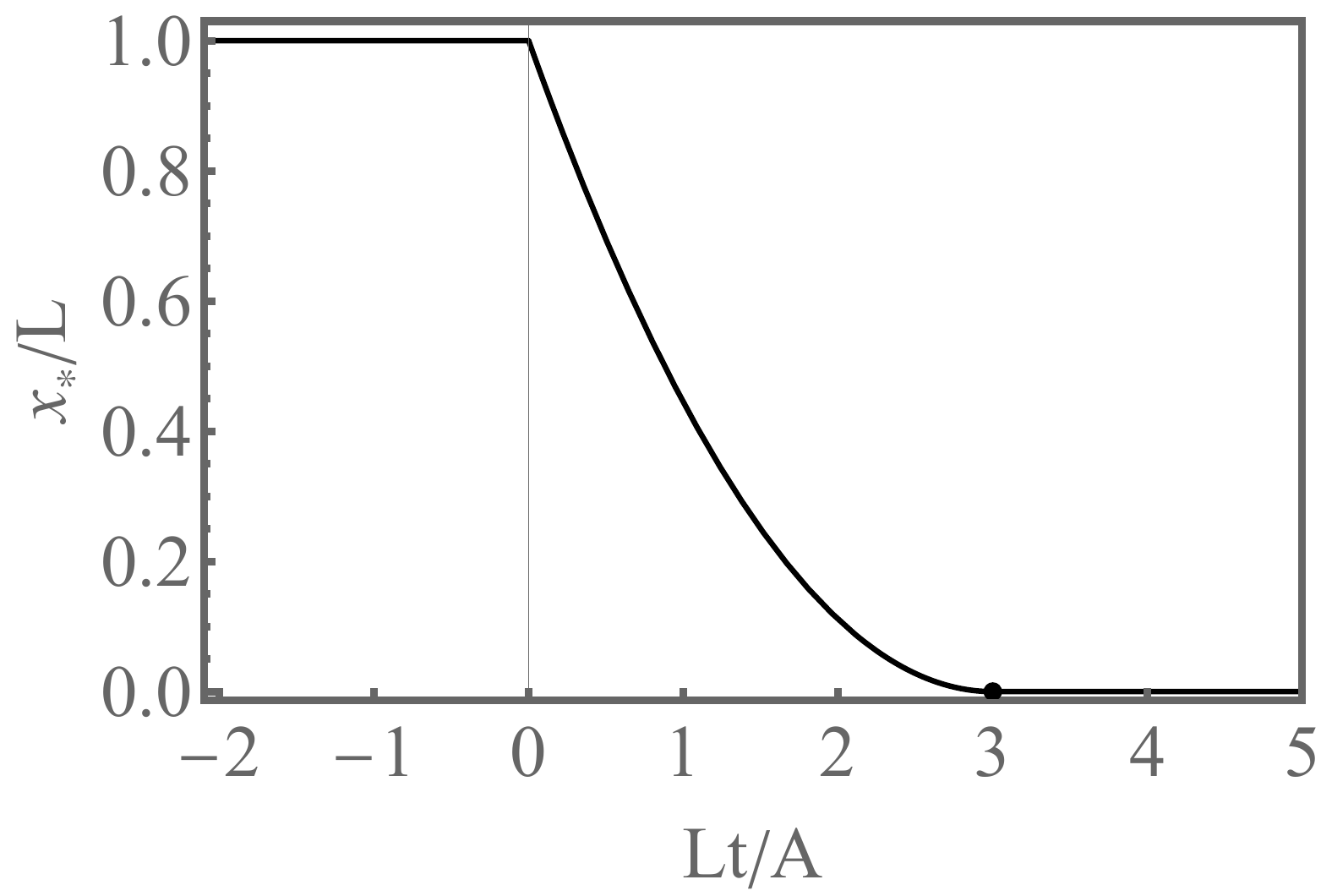}
\caption{Optimal paths dominating the fBm starting at $x=L$ and conditioned on a given area $A$ until the first passage time to $x=0$ for $H=1/4$ (the blue line) and $H=3/4$ (the magenta line). For comparison, the black line shows the optimal path for the standard Brownian motion, $H=1/2$, see Ref. \cite{MM2020a}.  The dots show the values of the optimal first passage time $T$.  The parameters are $A=L=1$. Here the geometrical optics can be made arbitrarily accurate by sending $D$ to zero.}
\label{fpareafig}
\end{figure}

\begin{figure}[ht]
\includegraphics[width=0.25\textwidth,clip=]{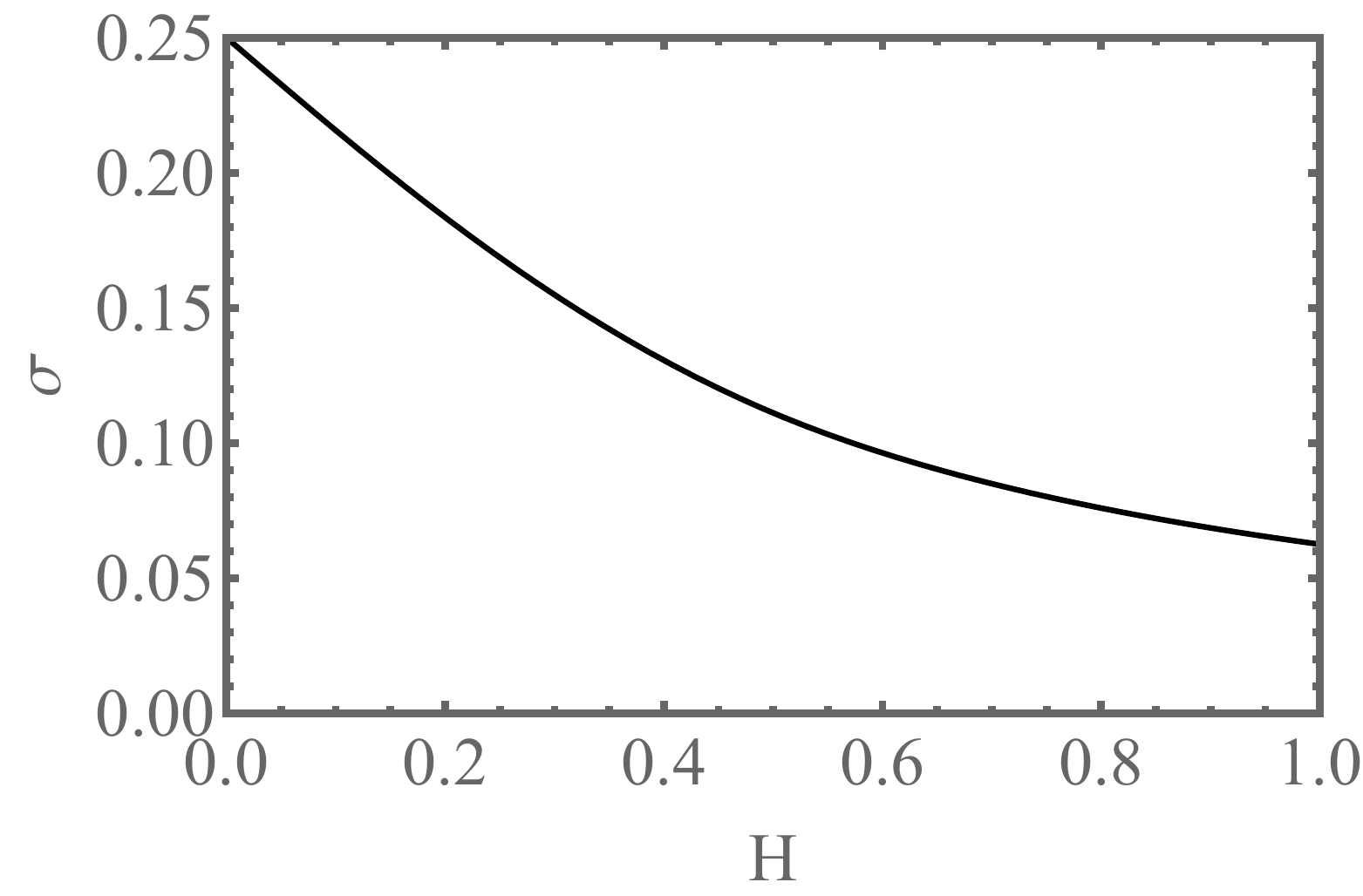}
\caption{The function $\sigma(H)$, see Eqs.~(\ref{SA})-(\ref{sigmamore}).}
\label{sigma}
\end{figure}

At a given $T$, the constrained action functional to be minimized coincides with that in Eq.~(\ref{functional2a}), leading to the same general solution (\ref{x200}). Now we demand that $x_*(T)=L$ and that the area (\ref{areadef}) be equal to $A$. These conditions yield
\begin{eqnarray}
\mu&=&\frac{2 (1+H) (2 A-L T)}{D (1-H)T^{2+2H}}\,, \nonumber \\
\lambda&=&\frac{2  A (1+H)-2 L T}{D (1-H) T^{1+2 H}} .
\label{mulambdafp}
\end{eqnarray}
At this stage our $x_*(t)$ still depends on $T$. The calculation of the action $S_T$ from Eq.~(\ref{actiongeneral}) at a given $T$  is almost identical to that in Sec. \ref{fBBfBE}, and we obtain
\begin{equation}\label{ST}
S_T(A,L) =\frac{2 A^2 (1+H)-2 A (1+H) L T+L^2 T^2}{2 D
   (1-H) T^{2+2H}}\,.
\end{equation}
Minimizing $S_T(A,L)$ with respect to $T$, we find that the minimum action is achieved at the optimal first-passage time $T=T(H)$, where
\begin{numcases}
{{T(H)} =}\!\frac{2A (1+H)}{L},& $H \leq 1/2$, \label{Tless} \\
 \!\frac{A(1+H)}{HL},& $H\geq 1/2$.
\label{Tmore}
\end{numcases}
The corresponding minimum action gives the $A \to 0$ tail of the probability distribution:
\begin{equation}\label{SA}
-\ln \mathcal{P}(A,L) \simeq \sigma(H) \frac{L^{2+2H}}{D A^{2H}}\,,
\end{equation}
where
\begin{numcases}
{{\sigma(H)} =}\!\!\!\frac{1}{2^{2+2H}(1-H^2)(1+H)^{2 H}},& \!\!$H \leq 1/2$ \label{sigmaless} \\
 \!\frac{H^{2H}}{2 (1+H)^{1+2 H}},& \!\!$H\geq 1/2$
\label{sigmamore}
\end{numcases}
Notable in Eq.~(\ref{SA}) is the essential singularity of $\mathcal{P}(A,L)$ at $A \to 0$, which character is determined by the Hurst exponent $H$. A graph of the function $\sigma(H)$ is shown in Fig. \ref{sigma}.

For the standard Brownian motion, $H=1/2$, we obtain $T(1/2) = 3A/L$, $\sigma(1/2)=1/9$, and
\begin{equation}\label{fphalf}
-\ln \mathcal{P}(A,L) \simeq \frac{L^{3}}{9 D A}
\end{equation}
in agreement with the exact result (\ref{PAexact}) and with the previous geometrical-optics result in this case \cite{MM2020a}.

Again, we will skip cumbersome explicit formulas for optimal path $x_*(t)$ which one obtains when
performing the integration in Eq.~(\ref{x200}) and using Eqs.~(\ref{mulambdafp}), (\ref{Tless}) and (\ref{Tmore}).  Three examples (after returning back to the original time $t$) are depicted in Fig. \ref{fpareafig}. At $H=1/2$ the non-trivial segment, $0<t<T$, of the optimal path is a parabola with a zero slope at $t=T=3A/L$ in agreement with the geometrical-optics result of Ref. \cite{MM2020a}. Again, at $H=1$ the optimal path becomes ballistic: in the original time variable $x_*(t) =L(1-Lt/2A)$, for all $|t|<\infty$.

\section{Summary and Discussion}
\label{discussion}

We showed that the geometrical optics makes it possible to efficiently evaluate large-deviation tails of a host of
statistics of the fBm. In addition to the distributions, the geometrical optics predicts optimal paths which provide valuable insights into the physics of fBm under ``unusual" constraints which push the fBm into large-deviation regimes.
One of our important findings is that the optimal paths of a fBm, subject to constraints on a finite interval $0<t\leq T$, also involve the past $-\infty<t<0$ and 
the future $T<t<\infty$. These features are natural consequences of the non-Markovian nature of the fBm. They do not violate causality, because the process is completely defined by its \textit{a priori} known statistics.

All our calculations were done for the two-sided fBm, $-\infty<t<\infty$.  They, however, can be immediately extended to the one-sided fBM, $0<t<\infty$, see Eq.~(\ref{kappa1}). 
The probability cost of a realization $x(t)$ in this case is
\begin{equation}\label{actiongeneral1}
S[x(t)] = \frac{1}{2} \int_{0}^{\infty} dt \int_{0}^{\infty} dt' K_1(t,t') x(t) x(t')\,,
\end{equation}
and the inverse kernel $K_1$ is defined by the relation
\begin{equation}\label{inverse1}
\int_{0}^{\infty} d \tau \,K_1(t, \tau) \,\kappa_1(\tau,t') =\delta(t-t'),
\end{equation}
and it differs from the kernel of the two-sided process \cite{a}. In spite of this difference,
the optimal paths of conditioned one-sided fBMs coincide, at $t>0$, with those
we found in this work. Furthermore, all the probability distribution tails that we evaluated here, remain exactly the same as for the two-sided process. Technically, this remarkable outcome is due to the delta-function(s) which emerge by virtue of the relation (\ref{inverse1}) and which are always localized on the studied interval $0<t\leq T$. 

It is useful to compare the geometrical-optics approach to the fBm with another perturbative approach \cite{Wieseetal,Wiese2019,SadhuWiese} which utilizes the proximity of the Hurst exponent $H$ to $1/2$ as a small parameter. The latter approach is not limited to large deviations. On the other hand, the geometrical-optics approach is uniformly valid for all $0<H<1$.

It would be very interesting to extend the geometrical-optics formalism to higher spatial dimensions and address  more complicated settings and geometries, such as encountered in a living cell and in other systems where anomalous diffusion is observed.  Now, after having tested the formalism on somewhat idealized physical problems, presented here, we can make further steps.

\section*{Acknowledgments}

We acknowledge a useful discussion with Pavel V. Sasorov. B.M. was supported by the Israel Science Foundation (Grant No. 1499/20) and by a
Chateaubriand fellowship of the French Embassy in Israel. He is very grateful to the
LPTMC, Sorbonne Universit\'{e}, for hospitality in the years of 2019-2020, when this work started.

\end{document}